\documentclass[twocolumn]{aastex62}

\usepackage{ulem}
\usepackage{amsmath}

\usepackage{color}

\begin{document}

\title{Double-Detonation Models for Type Ia Supernovae: Trigger of
  Detonation in Companion White Dwarfs and Signatures of Companions'
  Stripped-off Materials}

\correspondingauthor{Ataru Tanikawa}
\email{tanikawa@ea.c.u-tokyo.ac.jp}

\author{Ataru Tanikawa}
\affiliation{Department of Earth Science and Astronomy, College of
  Arts and Sciences, The University of Tokyo, 3-8-1 Komaba, Meguro-ku,
  Tokyo 153-8902, Japan; tanikawa@ea.c.u-tokyo.ac.jp}
\affiliation{RIKEN Center for Computational Science,
  7-1-26 Minatojima-minami-machi, Chuo-ku, Kobe, Hyogo 650-0047,
  Japan}

\author{Ken'ichi Nomoto}
\affiliation{Kavli Institute for the Physics and Mathematics of the
  Universe (WPI), The University of Tokyo, 5--1--5, Kashiwanoha,
  Kashiwa, 277--8583, Japan}

\author{Naohito Nakasato}
\affiliation{Department of Computer Science and Engineering,
  University of Aizu, Tsuruga Ikki-machi Aizu-Wakamatsu, Fukushima,
  965-8580, Japan}

\author{Keiichi Maeda}
\affiliation{Department of Astronomy, Kyoto University,
  Kitashirakawa-Oiwake-cho, Sakyo-ku, Kyoto 606-8502, Japan}

\begin{abstract}

We have studied double-detonation explosions in double-degenerate (DD)
systems with different companion white dwarfs (WD) for modeling type
Ia supernovae (SNe~Ia) by means of high-resolution smoothed particle
hydrodynamics (SPH) simulations. We have found that only the primary
WDs explode in some of the DD systems, while the explosions of the
primary WDs induce the explosions of the companion WDs in the other DD
systems. The former case is so-called Dynamically-Driven
Double-Degenerate Double-Detonation (D$^6$) explosion, or
helium-ignited violent merger explosion. The supernova ejecta of the
primary WDs strip materials from the companion WDs, whose mass is
$\sim 10^{-3}M_\odot$. The stripped materials contain carbon and
oxygen when the companion WDs are carbon-oxygen (CO) WDs with He
shells $\lesssim 0.04M_\odot$. Since they contribute to low-velocity
ejecta components as observationally interfered for iPTF14atg, D$^6$
explosions can be counterparts of sub-luminous SNe~Ia. The stripped
materials may contribute to low-velocity C seen in several SNe~Ia. In
the latter case, the companion WDs explode through He detonation if
they are He~WDs, and through double-detonation mechanism if they are
CO~WDs with He shells. We name these explosions ``triple'' and
``quadruple'' detonation (TD/QD) explosions after the number of
detonations. The QD explosion may be counterparts of luminous SNe~Ia,
such as SN~1991T and SN~1999aa, since they yield a large amount of
$^{56}$Ni, and their He-detonation products contribute to the early
emissions accompanying such luminous SNe~Ia. On the other hand, the TD
explosion may not yield a sufficient amount of $^{56}$Ni to explain
luminous SNe~Ia.

\end{abstract}


\section{Introduction}
\label{sec:introduction}

Type Ia supernovae (SNe~Ia) are one of the most luminous and common
explosive objects in the universe, and are utilized as a cosmic
standard candle. However, their progenitors have been unclarified so
far.  Although it is widely accepted that an SN~Ia is powered by a
white dwarf (WD) explosion \citep{1960ApJ...132..565H}, the nature of
its companion has been an open question. There are two popular
scenarios on the companion types. The first is a non-degenerate star
(main-sequence or red-giant stars), so-called single degenerate (SD)
scenario \citep{1973ApJ...186.1007W,1984ApJ...286..644N}. The second
possibility is another WD, namely double degenerate (DD) scenario
\citep{1984ApJS...54..335I,1984ApJ...277..355W}. The mass of an
exploding WD has been also under debate; it can be near-Chandrasekhar
mass \citep{1973ApJ...186.1007W}, or sub-Chandrasekhar mass
\citep{1982ApJ...257..780N}.

Recent observations have imposed several constraints on the nature of
their progenitors. It has been argued that the SD scenario may be
disfavored for some SNe~Ia. There are no red-giant stars in the
pre-explosion images of SN~2011fe and SN~2014J
\citep[][respectively]{2011Natur.480..348L,2014ApJ...790....3K}. A
supernova remnant LMC SNR 0509-67.5 has no surviving main-sequence
star \citep{2012Natur.481..164S,2017ApJ...837..111L}. However, we note
that spin-up/spin-down models can explain these non-detections
\citep{2011ApJ...730L..34J,2011ApJ...738L...1D,2012ApJ...756L...4H,2015ApJ...809L...6B}. On
the other hand, a large amount of circumstellar materials (CSMs) have
been indicated for PTF11kx \citep{2012Sci...337..942D}, which supports
the SD scenario. SNe~Ia resulting from the He detonation (usually
attributed for the sub-Chandrasekhar WDs) have been suggested, such as
MUSSES1604D \citep{2017Natur.550...80J}, and ZTF18aaqeas/SN~2018byg
\citep{2019arXiv190100874D}. X-ray observation with Hitomi telescope
has revealed that both of near- and sub-Chandrasekhar mass explosions
are required to explain the abundance pattern of the iron-peak
elements in the Perseus cluster \citep{2017Natur.551..478H}. There is
growing evidence that SNe~Ia can have multiple progenitors channels
\citep[e.g.][]{2013FrPhy...8..116H,2016IJMPD..2530024M}.

It is then important to make clear the dominant progenitor of the
standard candle and the origin of iron-peak elements. The
Dynamically-Driven Double-Degenerate Double-Detonation (D$^6$) model
(or helium-ignited violent merger model), one of sub-Chandrasekhar
mass models in the DD scenario, can be a promising candidate as the
dominant population of sub-Chandrasekhar mass explosion. The
discoveries of hypervelocity (HV) WDs \citep{2018ApJ...865...15S} have
strongly advocated the D$^6$ model, since the D$^6$ model results in a
WD thermonuclear explosion and a HV~WD remnant. The thermonuclear
explosion completely disrupts the primary WD. The companion WD
suddenly becomes free from the gravity of the primary WD, and is flung
away at a hypervelocity as a HV~WD. Although the current number of the
confirmed HV~WDs is too small for the D$^6$ model to explain all the
SNe~Ia, the observational sample may still be incomplete; an
increasing number of HV~WDs may be discovered in the future.

Since the D$^6$ model needs small amounts of He materials so as to
ignite He detonation by hydrodynamical effects, it leaves only a small
amount of He-detonation products; it is consistent with properties of
MUSSES1606D \citep{2017Natur.550...80J,2018ApJ...861...78M}, and
ZTF18aaqeas/SN~2018byg \citep{2019arXiv190100874D}.

The D$^6$ model is more advantageous than other sub-Chandrasekhar mass
explosion models in various respects. In the Milky Way galaxy, the
merger rate of DD systems with the super-Chandrasekhar mass in total
may be smaller than the SNe~Ia rate, although the total merger rate of
DD systems is comparable to the SNe~Ia rate
\citep{2014ARA&A..52..107M,2018MNRAS.476.2584M}. These arguments would
support the D$^6$ model, since for the D$^6$ model DD systems do not
necessarily exceed the Chandrasekhar mass in total. Note that other
sub-Chandrasekhar mass models, such as violent merger model
\citep{2010Natur.463...61P,2015ApJ...807..105S,2016ApJ...821...67S},
spiral instability model
\citep{2015ApJ...800L...7K,2017ApJ...840...16K}, and detached DD model
\citep{2016MNRAS.462.2486F}, requires DD systems with the
super-Chandrasekhar mass in total. Collisional DD models can explode
successfully even if their total mass is sub-Chandrasekhar mass
\citep{2009MNRAS.399L.156R,2009ApJ...705L.128R,2010MNRAS.406.2749L,2015MNRAS.454L..61D}. However,
the rate of such events have been unclear
\citep{2013MNRAS.430.2262H,2018A&A...610A..22T}.

The discovery of the HV~WDs has motivated us to examine the D$^6$
model in detail. We aim at revealing the properties of the D$^6$ model
as astronomical transients. In \cite{2018ApJ...868...90T} (hereafter,
Paper~I), we have reproduced the D$^6$ explosion of a DD system by
means of SPH simulation coupled with nuclear reaction networks, and
have investigated properties of its supernova (SN) ejecta and
surviving WD.  The main conclusions are summarized as follows. (1) The
SN ejecta have a velocity shift, $\gtrsim 1000$~km~s$^{-1}$, due to
the binary motion of the exploding WD. (2) The SN ejecta contain
low-velocity O components stripped from the companion WD, and the
stripped O may be observed in nebular phase spectra. However, since
the companion WD has no He shell, it was unclear if the O components
can be stripped from the companion WD when the companion WD has an
outer He shell. 

In this paper, we follow the D$^6$ explosions for various DD
systems. The D$^6$ explosions have distinct features from the violent
merger model, since they happen before the DD systems merge. Thus, the
companion WDs can survive the explosions of the primary WDs. We focus
on the interactions of the primary explosions with the companion
WDs. In Paper~I, we found O components will be stripped from the
companion WD when the companion WD has no He shell. At this time, we
assess whether the O components can be stripped from the companion WD
even when the companion WD has an outer He shell. Moreover, we
consider a He~WD or a CO~WD with an outer He shell as a companion WD,
and examine if the companion WD also explodes. We should remark
\cite{2015MNRAS.449..942P} have shown the explosion of the companion
He~WD, so-called triple-detonation (TD) explosion. In additional to
this explosion type, we find the explosion of the companion CO~WD,
which we call quadruple-detonation (QD) explosion. The TD/QD
explosions will have common features with the collisional DD models,
since both two WDs explode in these models.

This paper is structured as follows. We present our method in
section~\ref{sec:method}. We show our results in
section~\ref{sec:result}. We discuss our results and summarize our
paper in sections~\ref{sec:discussion} and \ref{sec:summary},
respectively.

\section{Method}
\label{sec:method}

Our SPH code is the same as used in Paper~I \citep[see
  also][]{2017ApJ...839...81T,2018ApJ...858...26T,2018MNRAS.475L..67T}. We
adopt the Helmholtz Equation of State \citep{2000ApJS..126..501T} for
equation of state, and the Aprox13 for nuclear reaction networks
\citep{2000ApJS..129..377T}. We parallelize our SPH code with the aid
of FDPS \citep{2016PASJ...68...54I}, and vectorize particle-particle
interactions by using the AVX instructions explicitly
\citep[e.g.][]{2012NewA...17...82T,2013NewA...19...74T}.

\begin{deluxetable*}{l|lllllll|llllllll}
  \tablecaption{Summary of initial
    conditions and simulation results. \label{tab:InitialCondition}}
  \tablehead{Model & $M_{\rm p}$ & $M_{\rm p,sh}$ & $M_{\rm p,He}$
    &$M_{\rm c}$ & $M_{\rm c,sh}$ & $r_{\rm sep,i}$ & $N$ & Exp. & $M_{\rm
      ej}$ & $M_{\rm ^{56}Ni}$ & $M_{\rm Si}$ & $M_{\rm O}$ & $M_{\rm
      cos}$ & $E_{\rm nuc}$ & $E_{\rm kin}$ \\
    & [$M_\odot$] & [$M_\odot$] & [$M_\odot$] & [$M_\odot$] & [$M_\odot$] & [$10^4$~km] & [$10^6$]
    &             & [$M_\odot$] & [$M_\odot$] & [$M_\odot$] & [$M_\odot$] & [$M_\odot$]
    & [Foe] & [Foe]
}
\startdata
He45R09  & $1.0$  & $0.05$ & $0.03$ & $0.45$ & --      & $2.9$ & $60$
         & TD
         & $1.45$ & $0.81$ & $0.15$ & $0.08$ & --
         & $2.3$  & $2.0$ \\
He45     & $1.0$  & $0.05$ & $0.03$ & $0.45$ & --      & $3.2$ & $60$
         & D$^6$
         & $0.98$ & $0.56$ & $0.15$ & $0.07$ & $0.0033$
         & $1.4$  & $1.1$ \\
CO60He00 & $1.0$  & $0.05$ & $0.03$ & $0.60$ & $0.000$ & $2.5$ & $67$
         & D$^6$
         & $0.97$ & $0.55$ & $0.15$ & $0.07$ & $0.0028$
         & $1.4$  & $1.1$ \\
CO60He06 & $1.0$  & $0.05$ & $0.03$ & $0.60$ & $0.006$ & $2.5$ & $67$
         & D$^6$
         & $0.97$ & $0.54$ & $0.15$ & $0.07$ & $0.0029$
         & $1.3$  & $1.1$\\
CO90He00 & $1.0$  & $0.10$ & $0.05$ & $0.90$ & $0.000$ & $1.6$ & $67$
         & D$^6$
         & $0.93$ & $0.51$ & $0.14$ & $0.06$ & $0.0024$
         & $1.4$  & $1.1$ \\
CO90He09 & $1.0$  & $0.10$ & $0.05$ & $0.90$ & $0.009$ & $1.6$ & $80$
         & D$^6$
         & $0.94$ & $0.52$ & $0.14$ & $0.06$ & $0.0033$
         & $1.4$  & $1.1$ \\
CO90He54 & $1.0$  & $0.10$ & $0.05$ & $0.90$ & $0.054$ & $1.6$ & $80$
         & QD
         & $1.90$ & $1.01$ & $0.28$ & $0.16$ & --
         & $2.5$  & $2.1$ \\
\enddata
\tablecomments{$M_{\rm p}$ and $M_{\rm p,sh}$ are the masses of the
  primary and primary's He shell, respectively, $M_{\rm p,He}$ is the
  He mass in the primary's He shell, and $M_{\rm c}$ and $M_{\rm
    c,sh}$ are the masses of companion and companion's He shell,
  respectively.  $r_{\rm sep,i}$ is the initial separation between the
  primary and companion. $N$ is the number of the SPH particles used
  for each model. ``Exp.'' means the explosion mode. $M_{\rm ej}$ is
  the total mass of the SN ejecta. $M_{\rm ^{56}Ni}$, $M_{\rm Si}$,
  and $M_{\rm O}$ are the masses of $^{56}$Ni, Si, and O in the SN
  ejecta, respectively. $M_{\rm cos}$ is the mass of the
  companion-origin stream. $E_{\rm nuc}$ is the nuclear energy
  released by the explosion, and $E_{\rm kin}$ is the kinetic energy
  of the SN ejecta.}
\end{deluxetable*}

We summarize our initial conditions in
Table~\ref{tab:InitialCondition}. We prepare $7$ DD systems. Although
all of them have super-Chandrasekhar mass in total, sub- and
super-Chandrasekhar systems should share some of common properties in
the D$^6$ explosions. The relaxation method to make single and binary
WDs is the same as in Paper~I. Every DD system has a $1.0M_\odot$
CO~WD as its primary WD. Each of the primary WD has a CO core and
outer He shell. These DD systems have various companion WDs: $2$
He~WDs with $0.45M_\odot$, $2$ CO~WDs with $0.60M_\odot$, and $3$
CO~WDs with $0.90M_\odot$. One of the $0.60M_\odot$ CO~WDs and two of
the $0.90M_\odot$ CO~WDs have outer He shells. The model names are
related to the total and He shell masses of the companion WDs.
CO60He00 is identical to the DD system in Paper~I. The DD systems have
circular orbits. On these orbits except for He45R09, the Roche-lobe
radii of the companion WDs are the same as their radii
\citep{1983ApJ...268..368E}. The semi-major axis in He45R09 is $0.9$
times that in He45. Although He45R09 should not appear in reality, we
prepare this system for the following two reasons. (1) We determine
the successful criteria of the TD explosion. (2) We investigate the
properties of the TD explosion which may succeed in a DD system with a
realistic separation if its primary WD has a different mass from
$1.0M_\odot$.

The CO~WDs (or their CO cores) consist of $50$~\% C and $50$~\% O by
mass, and the He~WDs consist of $100$~\% He. For the primary WDs,
their He shells are mixed with CO compositions composed of $50$~\% C
and $50$~\% O by mass. The 3rd and 4th columns indicate the total and
He mass in the He shell of each primary WD, respectively. Thus, the He
shells consist of $60$~\% ($50$~\%) He, $20$~\% ($25$~\%) C, and
$20$~\% ($25$~\%) O by mass for He45R09, He45, CO60He00, and CO60He06
(CO90He00, CO90He09, and CO90He54). For the companion WDs, the He
shells have pure He component.

Although the D$^6$ model can succeed even when the primary WD has a
thin He shell (say $0.01M_\odot$), we set up the primary WDs in all
the models with a thick He shell ($\sim 0.05M_\odot$) in order to
easily generate He and CO detonations in our SPH
simulations. Therefore, we should be careful of chemical elements
synthesized by He detonation. The primary WDs have their He shells
consisting of He and CO compositions. The mixing also makes the
initiation and propagation of He detonation easier
\citep{2014ApJ...797...46S}. The mixing can be achieved in DD systems
in reality due to Kelvin-Helmholtz instability in the merging process
of DD systems \citep{2013ApJ...770L...8P}. CO90He54 may have an
unrealistically thick He shell
\citep{1985ApJS...58..661I,1987ApJ...317..717I,1988ApJ...328..207K,1991ApJ...370..615I,2019MNRAS.482.1135Z}.
We set up this model to investigate features of the QD explosion.

\begin{figure}[ht!]
  \plotone{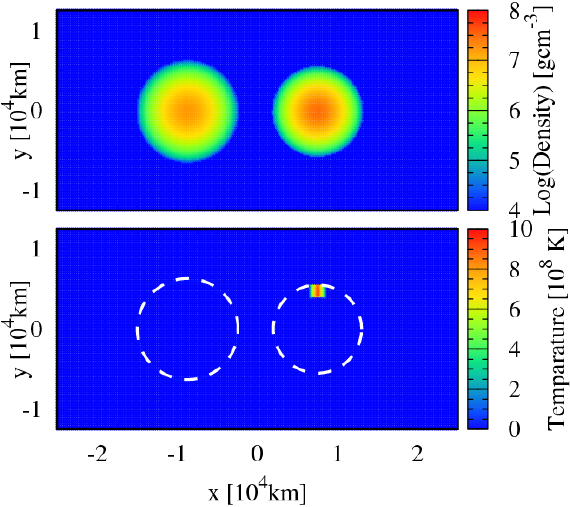}
  \caption{Initial density and temperature colormaps on the orbital
    plane for CO90He00. These are slices, not projections, which is
    the same in the other figures. For the temperature colormap, we
    delineate the primary and companion WDs by white dashed curves.}
  \label{fig:initCO90He00}
\end{figure}

As an example, we show the initial density and temperature
colormaps for CO90He00 in Figure~\ref{fig:initCO90He00}. The center of
mass of the DD system is located at the coordinate origin.  The
primary and companion WDs rotate counter-clockwise around each other
on the $x$-$y$ plane. Thus, the orbital angular momentum points to the
positive $z$-direction. In the primary WD, we set a hotspot large
enough to generate a He detonation in the He shell. The hotspot is at
the $x$-$y$ plane in the propagating direction of the primary WD, as
seen in the bottom panel of Figure~\ref{fig:initCO90He00}. We define
the beginning of the He detonation as $t=0$~s.

In all the models, an SPH particle has $m_{\rm SPH} \sim 2.4 \cdot
10^{-8}M_\odot$. This corresponds to mass resolution with $4.2$
million SPH particles per $0.1M_\odot$. The total numbers of SPH
particles are $60$ millions for He45R09 and He45, $67$ millions for
CO60He00 and CO60He06, and $80$ millions for CO90He00, CO90He09, and
CO90He54, summarized in Table~\ref{tab:InitialCondition}.  The SPH
kernel length $R$, which is effective space resolution, can be written
as
\begin{align}
  R \sim \left( \frac{m_{\rm SPH}}{\rho} \right)^{1/3} \sim 36 \left(
  \frac{\rho}{10^6~\mbox{g~cm$^{-3}$}} \right)^{-1/3} \;\;
  \mbox{[km]}, \label{eq:resolution}
\end{align}
where $\rho$ is density. Note that the SPH kernel length (and the
effective space resolution) in our SPH code becomes smaller for higher
density.

Although our simulations have high-mass resolution with $\sim 10^{-8}
M_\odot$, the resolution is not high enough to resolve detonation
waves. Thus, we may highly simplify the initiation of detonation, and
detonation itself. In appendix, we compare detonations in our
simulations with those in more realistic modelings.

In this paper, we use a polar coordinates system as well as the
Cartesian coordinates system to draw figures. We adopt the ordinary
way to transform the Cartesian coordinates system to the polar
coordinates system.

\section{Results}
\label{sec:result}

\subsection{Overview}
\label{sec:overview}

We summarize our simulation results as well as our initial conditions
in Table~\ref{tab:InitialCondition}. We can categorize the results of
our DD systems into three explosion types: D$^6$ explosion, TD
explosion, and QD explosion, as seen in the ``Exp.'' column. In the
D$^6$ explosion, just the primary WD explodes as expected in the D$^6$
model. The D$^6$ explosions occurs in He45, CO60He00, CO60He06,
CO90He00, and CO90He09. On the other hand, both the primary and
companion WDs explode in the cases of TD and QD explosions. DD systems
in He45R09 and CO90He54 experience the TD and QD explosions,
respectively. We describe in detail the D$^6$, TD, and QD explosions
in sections~\ref{sec:D6Explosions},
\ref{sec:TripleDetonationExplosion}, and
\ref{sec:QuadrupleDetonationExplosion}, respectively.

\subsection{D$^6$ Explosions}
\label{sec:D6Explosions}

Since we describe the D$^6$ explosion process in detail in Paper~I, we
briefly overview the explosion process here. In a DD system,
Roche-lobe overflow occurs from the lighter WD (companion WD) to the
heavier WD (primary WD). When the primary or companion WD has an outer
shell consisting of helium (He) materials, the He materials are
ignited due to hydrodynamical effects, and He detonation starts on the
primary surface \citep{2010ApJ...709L..64G,2013ApJ...770L...8P}. The
processes from the He detonation to carbon (C) ignition in the
carbon-oxygen (CO) core are the same as the double-detonation
mechanism. The He detonation does not ignite CO detonation in the CO
core of the primary WD directly, differently from the classical
double-detonation model
\citep{1982ApJ...257..780N,1986ApJ...301..601W}. The He detonation
surrounds the CO core. A shock wave separated from the He detonation
invades into the CO core, and converges at some point of the CO
core. The convergence of the shock wave ignites C burning, and
generates CO detonation
\citep{1990ApJ...354L..53L,1990ApJ...361..244L}.  The CO detonation
explodes the primary WD.  The explosion generates a blast wave. The
blast wave strips materials from the companion WD. The companion WD
conversely acts as an obstacle against the blast wave, and forms an
ejecta shadow.

In Table~\ref{tab:InitialCondition}, we see the masses of ejecta,
$^{56}$Ni, Si, and O for the D$^6$ explosions. The total ejecta mass
is about $1.0M_\odot$, which is the primary WD mass. However, it is
slightly less than $1.0M_\odot$. This is because small amounts of SN
ejecta are captured by the companion WDs. The captured masses are
$0.02M_\odot$ for He45, $0.03M_\odot$ for CO60He00 and CO60He06, and
$0.06$ - $0.07M_\odot$ for CO90He00 and
CO90He09. \cite{2017ApJ...834..180S} have analytically estimated that
the captured masses are $0.006$, $0.03$, and $0.08M_\odot$ for
companion WDs with $0.3$, $0.6$, and $0.9M_\odot$, respectively. Our
results are in good agreement with the Shen's estimate.

The $^{56}$Ni, Si, and O masses are similar among our D$^6$ models. As
described in Paper~I, these chemical abundances are consistent with
the results of previous studies
\citep[e.g.][]{2010A&A...514A..53F,2011ApJ...734...38W}.

We set thick He shells to generate He detonations easily.  However, it
has been discussed that the mass of the He shell could be smaller than
the value adopted here, to initiate the He detonation. Indeed, such a
scenario has been proposed for the D$^6$ explosion model to be
consistent with properties of normal SNe~Ia. The nucleosynthetic
yields for such a thin He shell condition could be approximated by the
results of the present model (thick He shell), as outlined
below. First, He detonation products are decreased. The He detonations
in our simulations yield heavier Si group elements (argon, calcium,
and titanium), not $^{56}$Ni, since the He shells consist of mixture
of He, C, and O. Thus, the amount of heavier Si group elements are
decreased by several $0.01M_\odot$, and the amount of $^{56}$Ni is
unchanged. Next, CO detonation products are increased. Since the He
shells occupy low-density regions, the CO detonation products should
be products of C and O burnings, i.e. O and lighter Si group elements
(silicon and sulfur). Consequently, the amount of O and lighter Si
group elements is increased by several $0.01M_\odot$. In summary, O,
Si, and S are increased by several $0.01M_\odot$ at the cost of the
decrease of heavier Si group elements. These chemical abundances are
still compatible to the results of the previous studies.

\begin{figure*}[ht!]
  \plotone{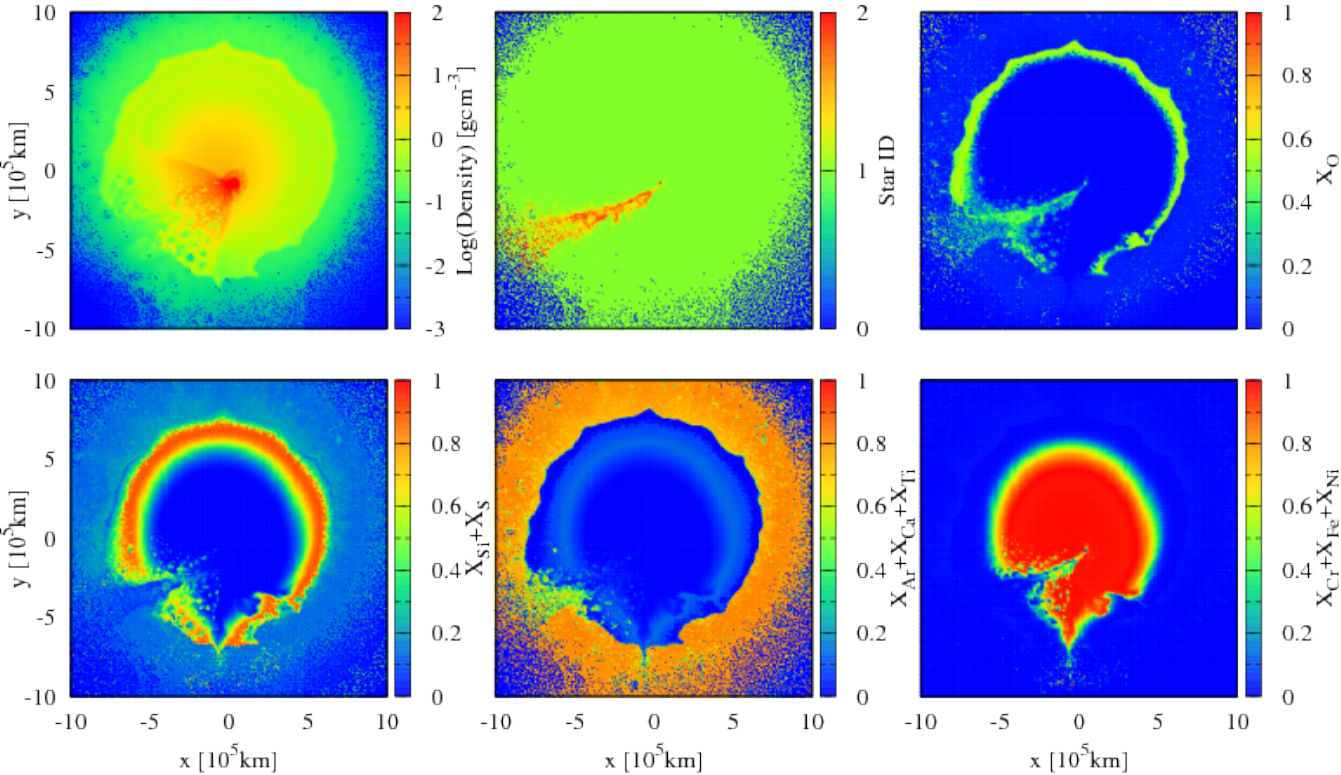}
  \caption{Density, star ID, and mass fraction of O, Si+S, Ar+Ca+Ti,
    and Cr+Fe+Ni at $t=50$~s for CO90He00. The star ID indicates the
    origin of materials. When the star IDs of a material are 1 and 2,
    the material originates from the primary and companion WDs,
    respectively.}
  \label{fig:snrCO90He00}
\end{figure*}

As an example, we show the SN ejecta at $t=50$~s in CO90He00 in
Figure~\ref{fig:snrCO90He00}. We can see an ejecta shadow
\citep{2015MNRAS.449..942P} in the density colormap. The colormap of
the star ID shows the presence of a companion-origin stream. The
companion-origin stream consists of CO as seen in the O colormap, and
does not contain He, since the companion WD has no He shell. Except
for the ejecta shadow and companion-origin stream, the SN ejecta has a
typical structure of the double-detonation explosion. Chemical
elements are distributed in a spherical form, and consist of
$^{56}$Ni, lighter Si group elements, O, and
heavier Si group elements from inside
to outside. The heavier Si group elements are yielded by the He
detonation.

\begin{figure*}[ht!]
  \plotone{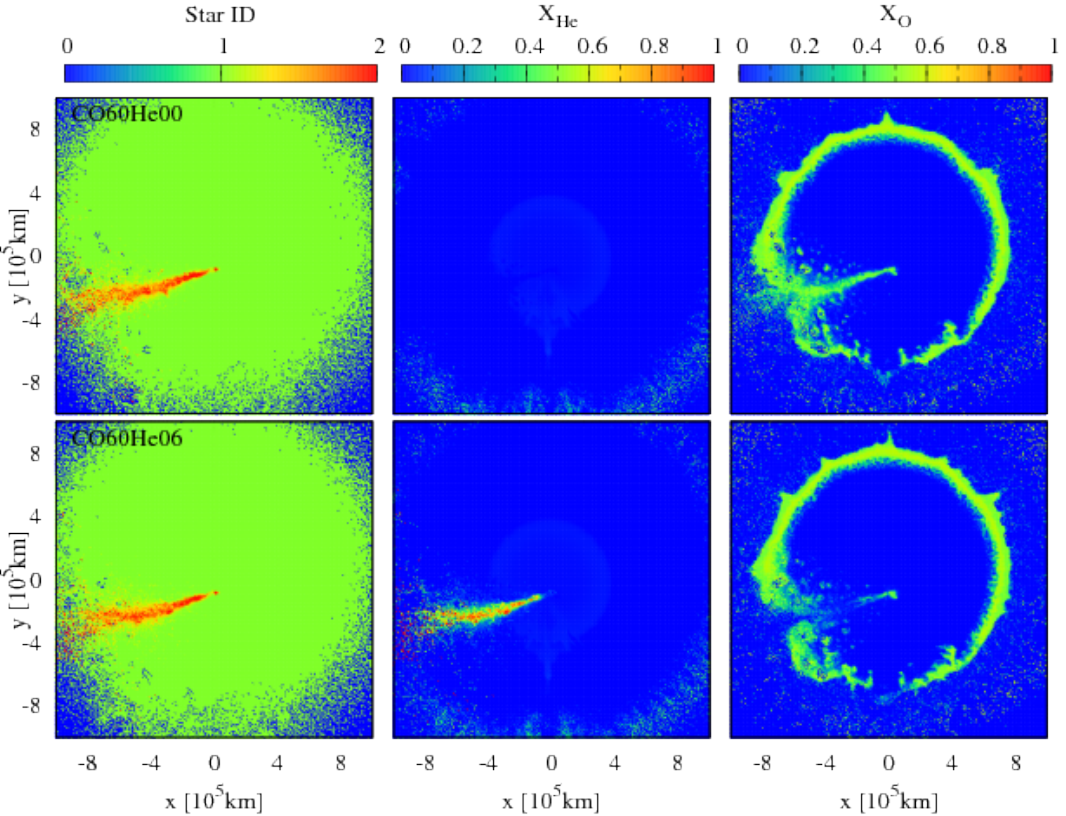}
  \caption{Star ID, and mass fraction of He and O at $t=50$~s for
    CO60He00 (top) and CO60He06 (bottom).}
  \label{fig:snrD6}
\end{figure*}

We investigate companion-origin streams. Figure~\ref{fig:snrD6} shows
star IDs, and mass fractions of He and O in CO60He00 and CO60He06 at
$t=50$~s. We can find from the left panels that both of the models
have companion-origin streams. CO60He00, which has no He shell in the
companion WD, does not contain He in their companion-origin streams,
while CO60He06, which has He shell in the companion WD, contains He in
their companion-origin streams (see the middle panels). For CO60He06,
its companion-origin stream also has O (see the right panels).

\begin{figure}[ht!]
  \plotone{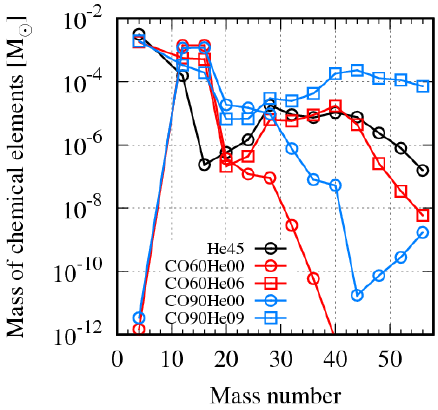}
  \caption{Chemical elements of companion-origin streams in He45,
    CO60He00, CO60He06, CO90He00, and CO90He09.}
  \label{fig:elementD6}
\end{figure}

Figure~\ref{fig:elementD6} shows chemical elements of companion-origin
streams in He45, CO60He00, CO60He06, CO90He00, and CO90He09. First, we
describe properties of models with CO~WD companions. The He masses in
CO60He00 and CO90He00 are much less than in CO60He06 and CO90He09. The
latter models have the companion-origin streams with large amounts of
unburned He. All the models have C+O in their companion-origin
streams. Nevertheless, the C+O masses are decreased with He shell
masses increasing. The masses of the lighter and heavier Si group
elements are slightly larger in models with He shells than in models
without He shells. He materials are burned by shock heating due to
collision of SN ejecta with companion WDs.

Next, we investigate properties of He45. The chemical elements in the
companion-origin stream are dominated by unburned He.  Slight amounts
of chemical elements other than He are present. These chemical
elements are also formed by shock heating through collision of the SN
ejecta with the companion WD.

\begin{figure}[ht!]
  \plotone{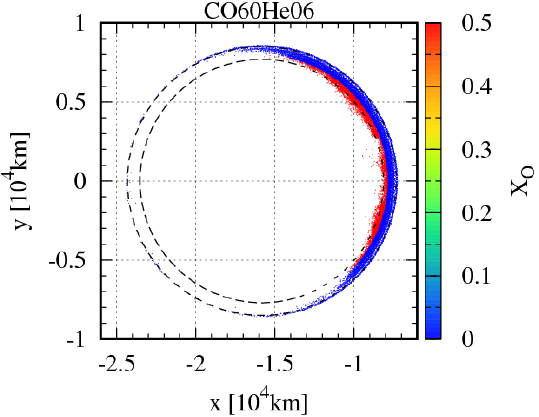}
  \caption{Initial position and oxygen mass fraction of stripped
    materials from the companion WD for CO60He06. We draw only
    materials with $|z| < 10^3$~km. The two dashed circles are used in
    Figure~\ref{fig:depthStrippedMass}. They are centered at the
    center of the companion WD. The outer one indicates the surface of
    the companion WD, and the inner one can have any radii.}
  \label{fig:strippedMass}
\end{figure}

The important point is that CO60He06 and CO90He09 as well as CO60He00
and CO90He00 include O materials in their companion-origin streams. In
particular, the presence of O in CO60He06 (and also CO90He09) is
unexpected in the following reason. Paper~I (and
Table~\ref{tab:InitialCondition}) shows the total mass of the
companion-origin stream is $2.8 \cdot 10^{-3}M_\odot$ in
CO60He00. Hence, if the outermost materials of CO60He06 is stripped,
the companion-origin stream should consist of He only. We investigate
the initial positions of materials in the companion-origin streams in
CO60He06, and draw their initial positions in
Figure~\ref{fig:strippedMass}. This figure clearly shows more
materials are stripped on the nearer side of the primary WD, as shown
analytically in \cite{1975ApJ...200..145W}, and numerically in
\cite{2014ApJ...792...66H}.  Therefore, O can be stripped on the
nearer side of the primary WD even if the companion WD has a He shell.

\begin{figure}[ht!]
  \plotone{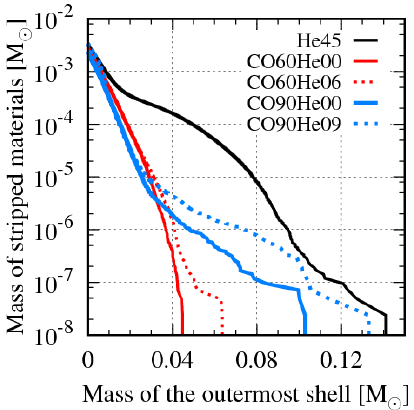}
  \caption{Mass of stripped materials inside of the inner dashed
    circle shown in Figure~\ref{fig:strippedMass}. The horizontal axis
    indicates the mass of a spherical shell between the two dashed
    circle in Figure~\ref{fig:strippedMass}. Note that the inner
    dashed circle can have any radii.}
  \label{fig:depthStrippedMass}
\end{figure}

Figure~\ref{fig:depthStrippedMass} shows the depth of the stripped
materials. As for models with companion CO~WDs, we find the depth
distributions of stripped materials are nearly independent of the
masses of He shells, comparing the depth between CO60He00 and
CO60He06, and between CO90He00 and CO90He09. Slightly more materials
are stripped with more massive He shell due to He burning. The reason
why the depth distribution in He45 is largely different from in other
models would be the intensity of He burning.

Using Figure~\ref{fig:depthStrippedMass}, we can conjecture the mass
of O in a companion-origin stream when a companion CO~WD has a He
shell. A companion-origin stream contains $\sim 10^{-4}$, $10^{-5}$
and $10^{-6}M_\odot$ of O, even if a companion WD has $0.013$,
$0.027$, and $0.04M_\odot$ of a He shell. When a companion WD has a He
shell with more than $0.04M_\odot$, the O mass is less than
$10^{-6}M_\odot$, and dependent on the companion mass.

\begin{figure*}[ht!]
  \plotone{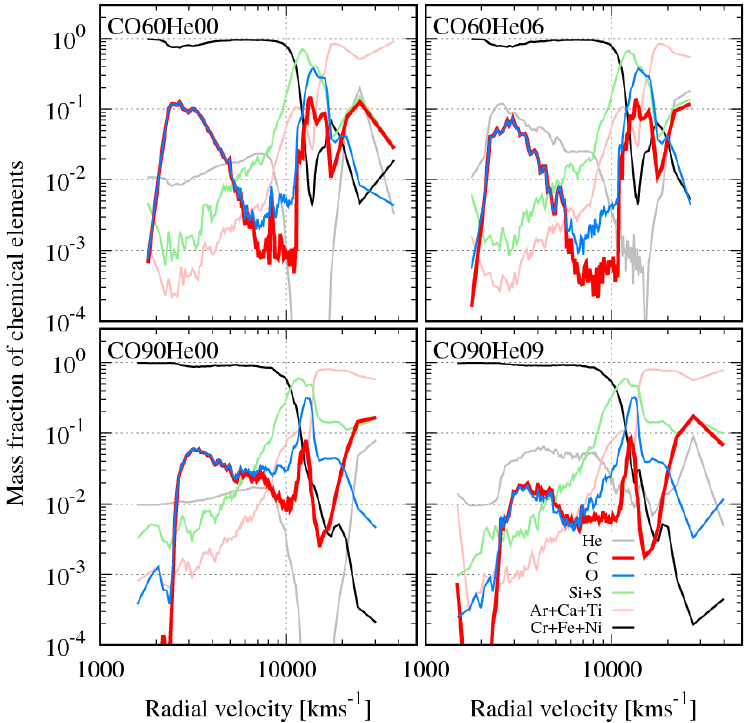}
  \caption{Mass fraction of chemical elements as a function of radial
    velocity at $t=50$~s from the viewpoint of $(\theta,
    \phi)=(90^\circ, 240^\circ)$ for CO60He00, CO60He06, CO90He00, and
    CO90He09.}
  \label{fig:angleD6}
\end{figure*}

Figure~\ref{fig:angleD6} shows mass fractions of chemical elements as
a function of radial velocity of SN ejecta from the viewing angle of
$(\theta, \phi)=(90^\circ, 240^\circ)$. From this viewing angle,
features of companion-origin streams look most prominent. In all the
models, the SN ejecta primarily have the heavier Si group elements, CO
components, the lighter Si group elements, $^{56}$Ni, and CO+He
components from higher velocity to lower velocity. Except for the
lower velocity CO+He components, the distribution of the chemical
elements are typical of the double-detonation explosion. The lower
velocity CO+He components come from the companion-origin
streams. Although the lower velocity CO+He components have less O with
He shell masses increasing, low-velocity O components are still
present. In future, we should assess whether such low-velocity O
components may be detected in the nebular phase of an SN~Ia.

In Paper~I, we found unburned materials due to numerical artifacts,
and convert them to $^{56}$Ni. In this paper, we leave them as they
are. They are captured by the companion WDs, and little is contained
in the SN ejecta. Since they are initially located at the centers of
the primary WDs, they get small velocities through the explosions of
the primary WD. Eventually, they do not contaminate the SN ejecta.

\subsection{Triple-Detonation (TD) Explosion}
\label{sec:TripleDetonationExplosion}

\begin{figure*}[ht!]
  \plotone{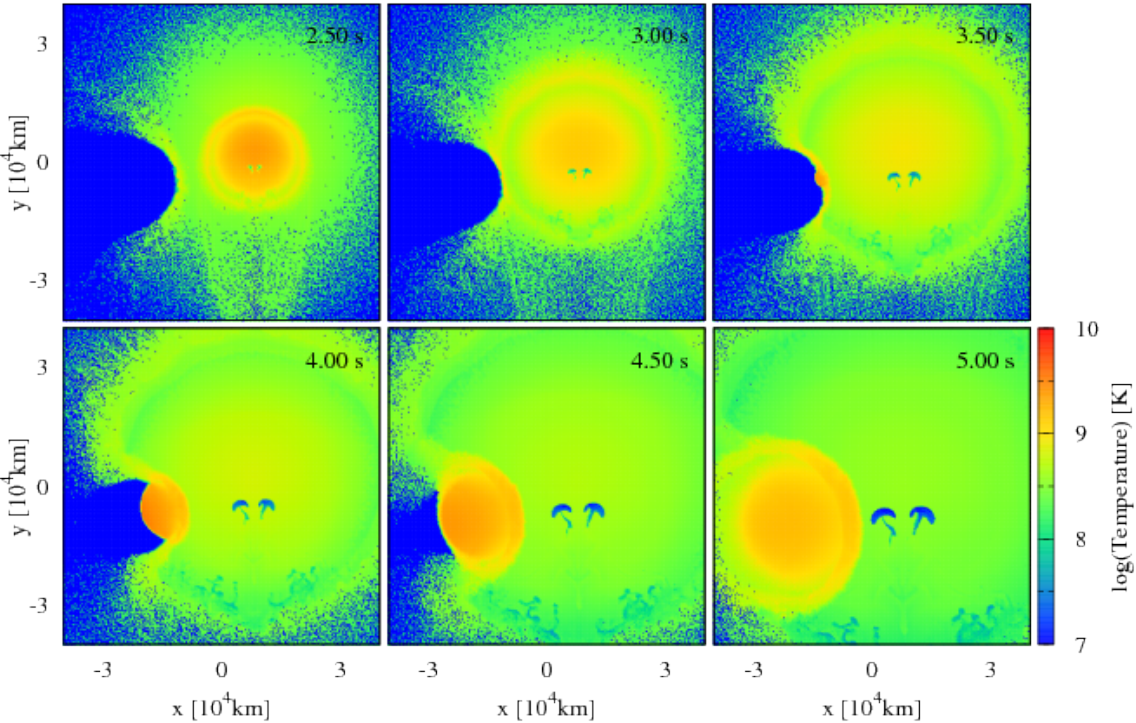}
  \caption{Temperature evolution for He45R09.}
  \label{fig:evolveTd}
\end{figure*}

In this section, we describe properties of SN ejecta of a TD explosion
which emerges in He45R09. Figure~\ref{fig:evolveTd} draws the
temperature evolution for He45R09. The processes are as follows. The
CO detonation disrupts entirely the primary WD by $t=2.50$~s. The
resultant blast wave ignites a He detonation in the companion He~WD
just before $t=3.50$~s. The He detonation disrupts the companion He~WD
by $t=5.00$~s.

A TD explosion occurs in He45R09, while it does not in He45. The
former binary separation is $2.9 \cdot 10^4$~km (or $0.041R_\odot$),
and the latter is $3.2 \cdot 10^4$~km (or $0.046R_\odot$). This is
quite consistent with the argument that a TD explosion is feasible if
a binary separation is $\lesssim 0.045R_\odot$ in
\cite{2015MNRAS.449..942P}. 

We should bear in mind that this DD system has an impractical
configuration. Nevertheless, we investigate features of this
explosion. This is because the TD explosion could occur if the primary
WD has a different mass from $1.0M_\odot$. For example, a DD system
with $0.90M_\odot$ and $0.45M_\odot$ WDs can be close to $\sim
0.045R_\odot$.

We can see mushroom-shaped unburned regions at the center of the
primary WD. We regard these unburned regions as numerical artifacts in
the same reason described in Paper~I. We do not suspect that these
unburned materials involve the beginning of the TD. As seen in the
panel at $t=3.50$~s, the TD explosion starts before these unburned
materials hit the companion WD.

The total ejecta mass is $1.45M_\odot$ (see
Table~\ref{tab:InitialCondition}), the same as the total mass of the
DD system. It is larger than those in the D$^6$ explosion models by
$\sim 0.5M_\odot$. The increment of the ejecta mass consists of
$^{56}$Ni and He. The $^{56}$Ni is the product of the He detonation in
the companion WD. The Si, O and C masses are similar to those in the
D$^6$ explosion model.

\begin{figure*}[ht!]
  \plotone{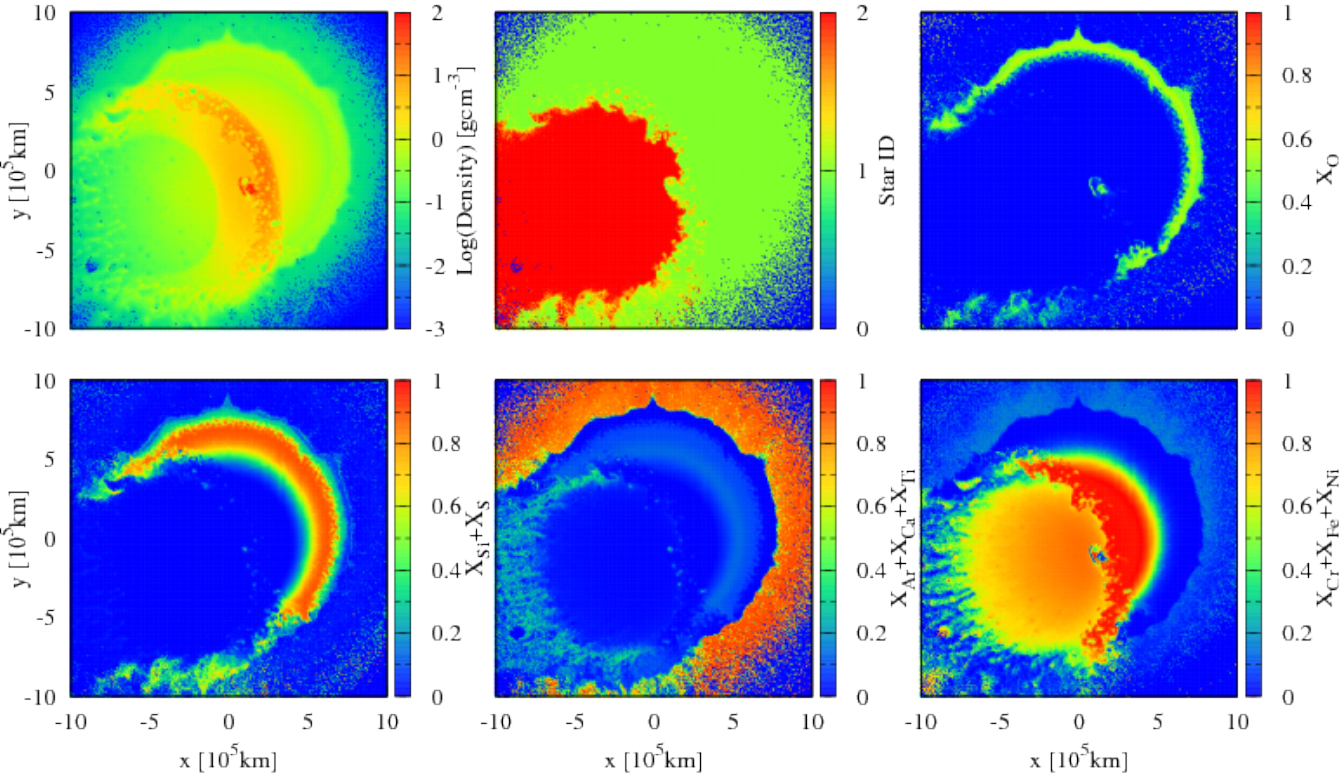}
  \caption{The same as in Figure~\ref{fig:snrCO90He00}, except for
    He45R09.}
  \label{fig:snrHe45R09}
\end{figure*}

Figure~\ref{fig:snrHe45R09} shows SN ejecta in He45R09 at
$t=50$~s. The SN ejecta have overlapping structure of the two
explosions. Since the companion WD explodes after the primary WD, the
companion explosion overlays the primary explosion. The density
colormap indicates a crescent-shaped, high density region. This region
is formed through pushing back of the companion explosion against the
primary explosion.

\begin{figure*}[ht!]
  \plotone{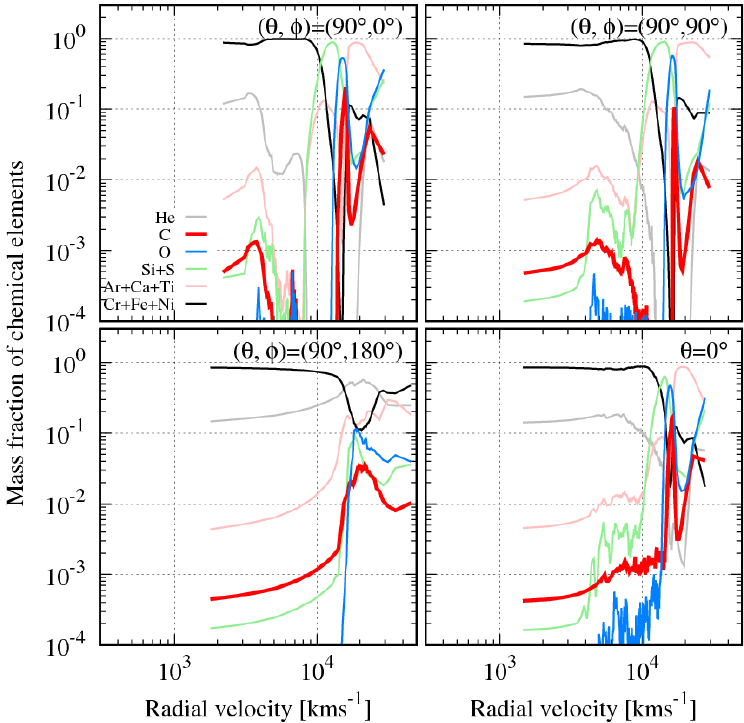}
  \caption{Mass fraction of chemical elements as a function of radial
    velocity at $t=50$~s for He45R09. The view point is indicated on
    the top-right side of each panel.}
  \label{fig:angleTd}
\end{figure*}

We show the mass fraction of chemical elements as a function of radial
velocity at $t=50$~s for He45R09 in Figure~\ref{fig:angleTd}. From all
the view points, low-velocity components ($\lesssim 10^4$~km) consist
mostly of $^{56}$Ni and partly of He ($\sim 10$~\%). This He materials
are embers of the He detonation in the companion WD. From all the view
points but $(\theta, \phi)=(90^\circ, 180^\circ)$, high-velocity
components ($\gtrsim 10^4$~km) are dominated by the lighter Si group
elements, C+O, and the heavier Si group elements from lower velocity
to higher velocity. This results from the primary's explosion, a
typical feature of double-detonation explosions. Chemical abundance
seen from the view point of $(\theta, \phi)=(90^\circ, 180^\circ)$ is
completely different those seen from the other view points. Even in
high-velocity components ($\gtrsim 10^4$~km), the dominant chemical
elements are He and $^{56}$Ni from lower velocity to higher
velocity. This is because we see directly the companion's explosion
from this view point.

\subsection{Quadruple-Detonation (QD) Explosion}
\label{sec:QuadrupleDetonationExplosion}

\begin{figure*}[ht!]
  \plotone{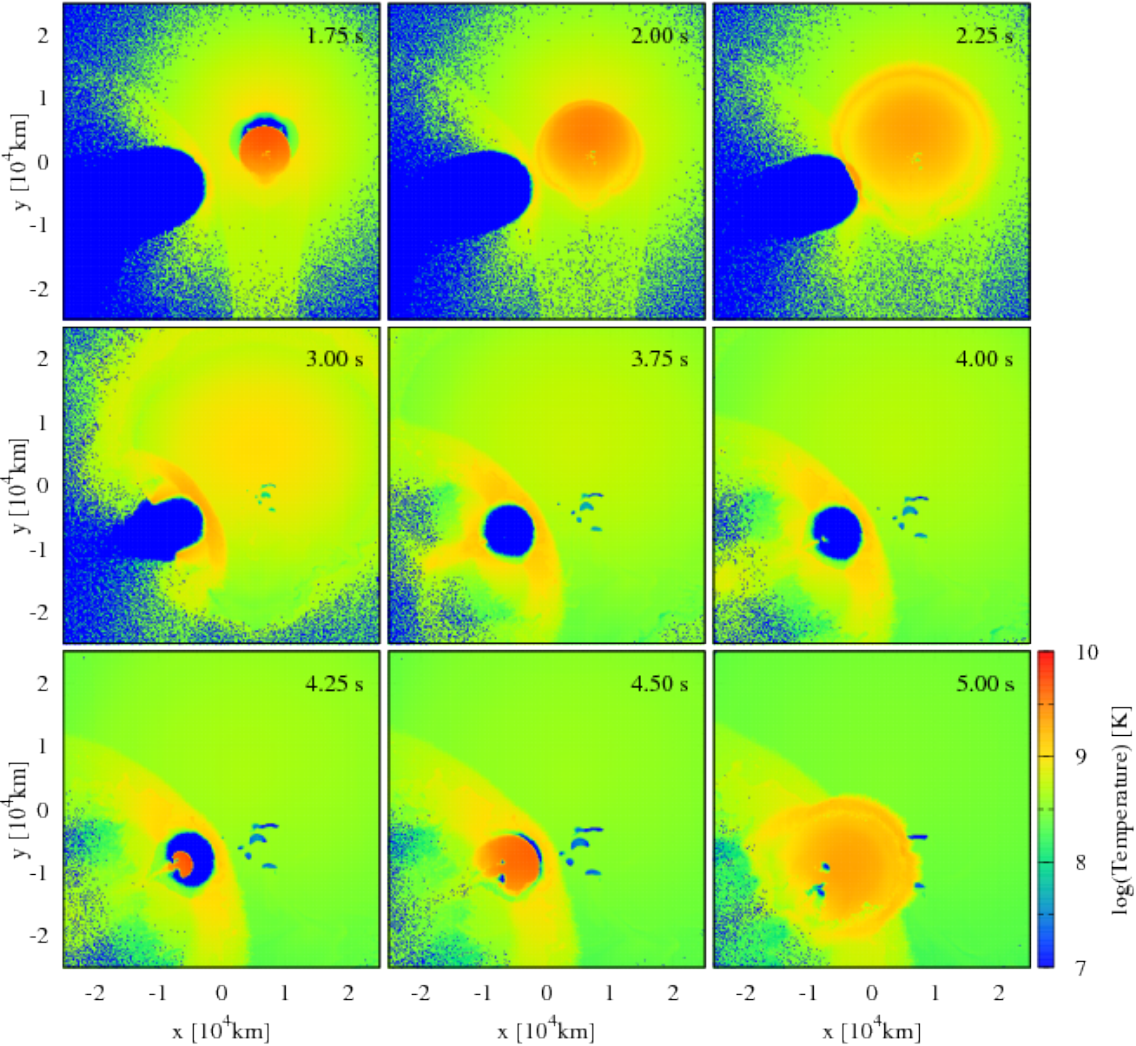}
  \caption{Temperature evolution for CO90He54.}
  \label{fig:evolveQd}
\end{figure*}

In this section, we show properties of SN ejecta of a QD
explosion. First, we present processes of the QD explosion for
CO90He54 in Figure~\ref{fig:evolveQd}. The CO detonation disrupts the
primary WD by $t=2.00$~s. The blast wave hits the companion WD, and
generates the He detonation in the companion WD at $t=2.25$~s. The He
detonation surrounds the companion WD by $t=3.75$~s, and sends a
converging shock wave into the companion WD. The converging shock
ignites the CO detonation in the CO core of the companion WD at
$t=4.00$~s. The CO detonation explodes the companion WD by $t=5.00$~s.

No QD explosion occurs in CO90He09. This is because the blast wave of
the primary's explosion completely strips the He shell of the
companion WD on the near side of the primary WD before the blast wave
ignites the He detonation.

Both the explosions of the primary and companion WDs leave unburned
materials at their centers due to numerical artifacts. We do not expect
these unburned materials ignite the QD for the same reason as the TD
case. The unburned materials have not yet collided with the companion
WD when the He and CO detonations in the companion WD start. When we
investigate observational features of the QD, we have to be careful of
the unburned materials from the companion WD. However, we need not be
careful of the unburned materials from the primary WD. This is because
the unburned materials in the primary WD are mixed with physically
unburned materials in the companion WD due to low density, and because
the former mass is much smaller than the latter mass.

\begin{figure*}[ht!]
  \plotone{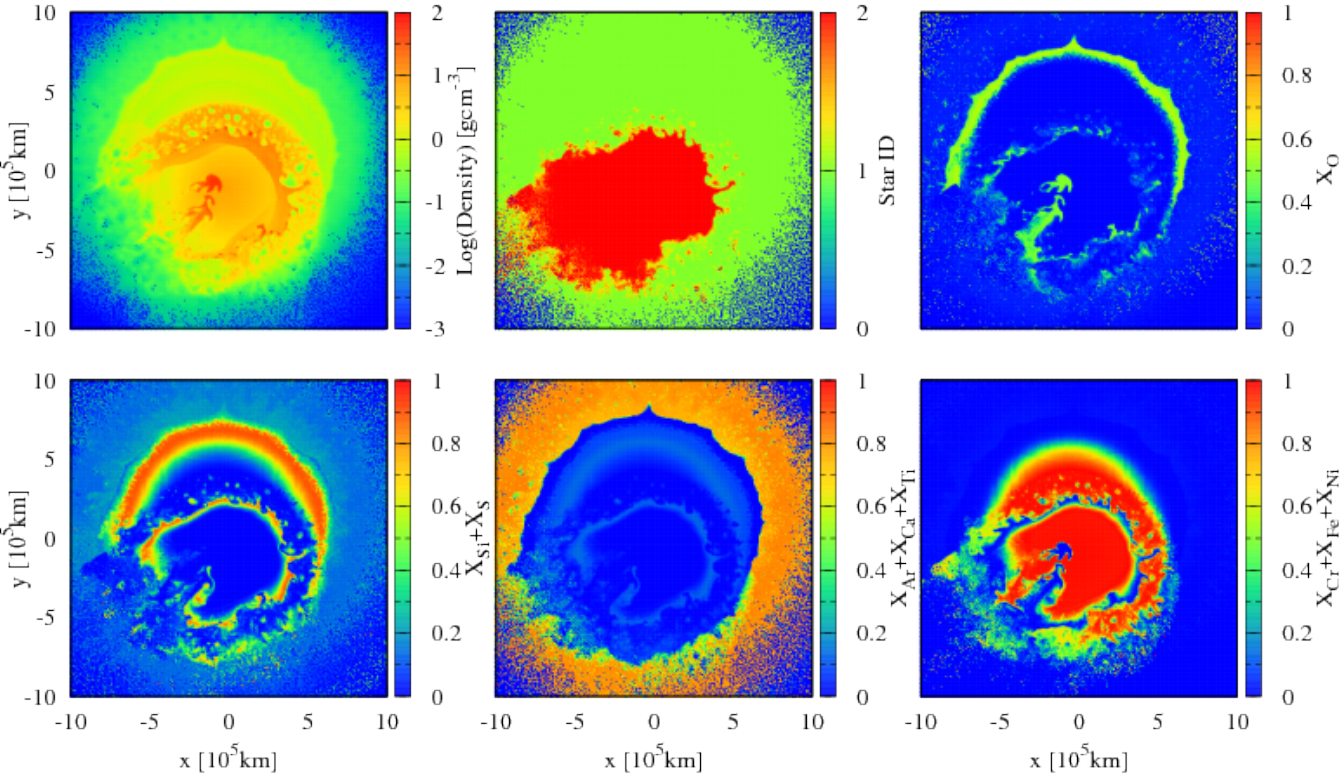}
  \caption{The same as in Figure~\ref{fig:snrCO90He00}, except for
    CO90He54.}
  \label{fig:snrCO90He54}
\end{figure*}

Figure~\ref{fig:snrCO90He54} shows the SN ejecta at $t=50$~s for
CO90He54. The SN ejecta have nested structure consisting of the
explosions of the primary and companion WDs. As seen in the star ID
panel, the outer and inner parts come from the primary and companion
WDs, respectively, since the primary WD explodes before. The QD
explosion have the nested structure, not overlapping structure like
the TD explosion. Since the QD explosion time is later than the TD
explosion time (see Figures~\ref{fig:evolveTd} and
\ref{fig:evolveQd}), the primary's SN ejecta in the QD explosion have
more time traveling farther away than in the TD explosion. From the
outside to the inside, the SN ejecta have the heavier Si group
elements, O, the lighter Si group elements, $^{56}$Ni, O, the lighter
Si group elements, $^{56}$Ni, and O. The heaver Si group elements in
the outermost of the SN ejecta are synthesized by the He detonation in
the primary WD. The outer $^{56}$Ni elements are composed of those
synthesized by the CO detonation in the primary WD, and by the He
detonation in the companion WD from the outside to the inside. The He
detonation in the companion WD mainly yields $^{56}$Ni, which is
different from the He detonation in the primary WD. This is because
the He shell of the companion WD consists of pure He.

The innermost part of the SN ejecta has O components, which are
unburned materials due to numerical artifacts. The mass of these
unburned materials is not large, $\sim 0.01M_\odot$. However, we have
to be careful that they contaminate chemical abundance in the inner
part of the SN ejecta.

\begin{figure*}[ht!]
  \plotone{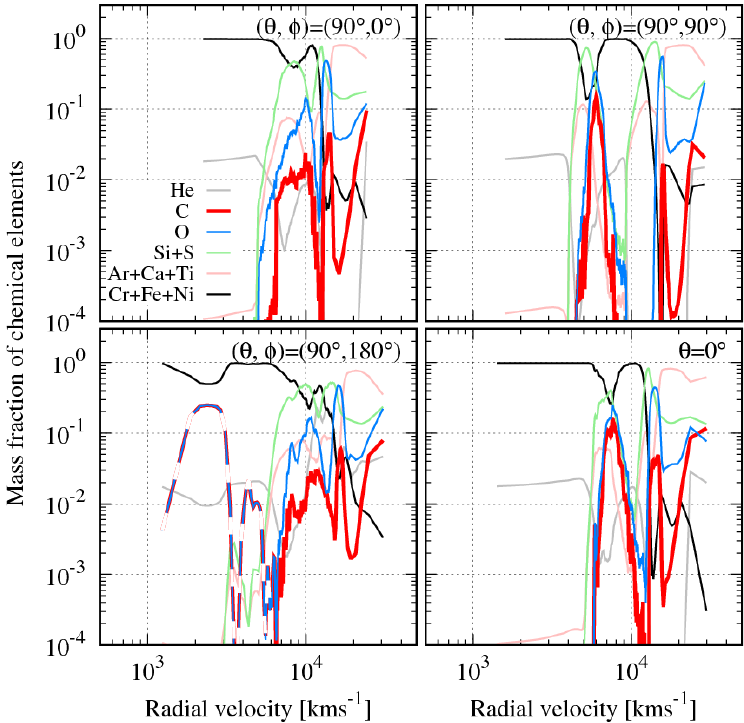}
  \caption{The same as in Figure~\ref{fig:angleTd}, except for
    CO90He54. In the panel of $(\theta,\phi)=(90^\circ,180^\circ)$,
    low-velocity CO with $< 6 \cdot 10^3$~km~s$^{-1}$ are indicated by
    dashed curves, since these components come from unburned materials
    due to numerical artifacts.}
  \label{fig:angleQd}
\end{figure*}

Figure~\ref{fig:angleQd} shows the chemical mass fractions as a
function of radial velocity from various viewing angles. As expected,
we can see the nested structure of the QD (two double-detonation)
explosions. From higher velocity to lower velocity, the SN ejecta
consist of He-detonation products, O, the lighter Si group elements,
$^{56}$Ni, O, the lighter Si group elements, and $^{56}$Ni. In panel
with $(\theta, \phi)=(90^\circ, 180^\circ)$, there are O components
with $\sim 2000$ - $3000$~km~s$^{-1}$, which come from unburned
materials due to numerical artifacts.

\section{Discussion}
\label{sec:discussion}

\subsection{Counterparts of D$^6$, TD and QD Explosions}

We discuss counterparts of D$^6$, TD, and QD explosions. As described
in Paper~I, the D$^6$ explosion may be consistent with sub-luminous
SN~Ia iPTF14atg. We will reproduce sub-luminous SNe~Ia, if we choose
masses of primary WDs to be less than $1.0M_\odot$. Moreover,
iPTF14atg has oxygen emission in its nebular phase
\citep{2016MNRAS.459.4428K}, which could be explained by a
companion-origin stream in D$^6$ explosion. As shown in the previous
section, a companion-origin stream can contain oxygen, even if a
companion WD has a He shell. However, we note that the amount of
oxygen is quite small. We should perform radiative transfer
calculation in order to assess whether oxygen in a companion-origin
stream can be detected in late time spectra of SNe~Ia.

Since the companion-origin stream also contains C, it may correspond
to low-velocity ($\lesssim 10000$~km~s$^{-1}$) C
\citep{2009ApJ...707L.118Y,2012ApJ...745...74F,2012MNRAS.425.1917S,2014ApJ...789...89C,2015A&A...578A...9H}.
Although C velocity is $\sim 3000$~km~s$^{-1}$ in
Figure~\ref{fig:angleD6}, the velocity may be $\sim 10000$~km~s$^{-1}$
from different views. This is because the companion-origin stream
spreads from the inner part of the SN ejecta to the outer part (see
the right panels in Figure~\ref{fig:elementD6}.

Since TD and QD explosions yield a large amount of $^{56}$Ni, we
expect their counterparts are super-Chandrasekhar SNe~Ia, or luminous
SNe~Ia, such as SN~1991T-like and SN~1999aa-like SNe~Ia. However, we
note that the TD explosion may not yield such a large amount of
$^{56}$Ni. The TD explosion may be achieved only when the primary WD
has less than $1.0M_\odot$ (more likely less than $0.9M_\odot$). In
this case, the primary WD yields $\sim 0.3M_\odot$ of $^{56}$Ni
\citep[e.g.][]{2018ApJ...854...52S}. Then, the total $^{56}$Ni mass
should be $\lesssim 0.6M_\odot$, which is not in agreement with those
of luminous SNe~Ia.

We thus discuss only the QD explosion. It may be difficult for the QD
explosion to explain super-Chandrasekhar
SNe~Ia. \cite{2016PASJ...68...68Y} have reported that SN~2012dn, a
candidate of super-Chandrasekhar SNe~Ia, has massive CSMs. Such
massive CSMs cannot be formed prior to the QD explosion.

We then discuss applicability of the QD explosion model to
SN~1991T/1999aa-like SNe~Ia. These luminous SNe~Ia have been said to
involve early excess optical/UV emission few days after their
explosion \citep{2018ApJ...865..149J}. This early excess emission
could be consistent with surface $^{56}$Ni formed by the He detonation
on the primary or companion WD. \cite{2018ApJ...864L..35S} have shown
that the early excess emission colors of luminous SNe~Ia are
relatively blue, $B-V \sim -0.1$. The He-detonation products can
explain such a color \citep{2018ApJ...861...78M}. Moreover, the
He-detonation products can consistently make the UV emission of
SN~1991T/1999aa-like SNe~Ia, such as SN~2012cg and SN~2017cbv
\citep{2018ApJ...861...78M}, while interactions with a non-degenerate
companion and CSMs may overproduce UV emission at the early phase
\citep[][respectively]{2017ApJ...845L..11H,2018ApJ...861...78M}.

As seen in Figure~\ref{fig:angleQd}, the QD explosion has high- and
low-velocity components of $^{56}$Ni. This may explain two-peak
emission lines of Co~III in SN~2007on
\citep{2018MNRAS.479L..70D}. Note that \cite{2018MNRAS.479L..70D} have
attributed such emission lines to the collisional DD models. Although
SN~2007on is not luminous SNe~Ia, such QD explosions can synthesize
$<0.6M_\odot$ of $^{56}$Ni mass if the primary and companion WDs have
$\lesssim 0.9M_\odot$. However, the QD explosion may not make such
prominent double-peak features of Co emission lines, since its
$^{56}$Ni distribution is nearly spherical, differently from $^{56}$Ni
distribution of the collisional DD models.

Finally, we mention a TD explosion. A TD explosion does not yield
$^{56}$Ni large enough to explain luminous SNe~Ia. However, its
spectra may be consistent with those of luminous SNe~Ia, if it is
observed from the side of the companion WD:
$(\theta,\phi)=(90^\circ,180^\circ)$ in Figure~\ref{fig:angleTd}. This
is because Si emission lines cannot be observed from this side, which
is consistent with luminous SNe~Ia.

\subsection{Comparison with Other Explosion Models}

We compare D$^6$, TD, and QD explosions with other sub-Chandrasekhar
mass explosions in DD systems, violent merger model
\citep{2010Natur.463...61P}, collisional DD model
\citep{2015MNRAS.454L..61D}, spiral instability model
\citep{2015ApJ...800L...7K}, and detached DD model
\citep{2016MNRAS.462.2486F}.

We can distinguish D$^6$ explosions from the collisional DD model by
oxygen emission lines in nebular phases. This model has no
low-velocity oxygen. On the other hand, the violent merger and spiral
instability models can have oxygen emission lines in the nebular phase
\citep{2013ApJ...775L..43T,2016MNRAS.459.4428K,2016ApJ...827..128V}. Thus,
we should identify D$^6$ explosions from the violent merger and spiral
instability models by velocity shift of supernova ejecta of D$^6$
explosions (Paper~I). This velocity shift comes from the binary motion
of the exploding WD, and is comparable to the velocity of a HV~WD,
$\sim 10^3$~km~s$^{-1}$. The difference between the D$^6$ and detached
DD model should be whether He detonation products are present or not.

We compare TD/QD explosions with the other DD models. The TD/QD
explosions could leave He detonation ashes. Thus, we can identify
these explosions from the other DD models with footprints of He
detonation products. If these explosions can have double-peak features
of Co emission lines in nebular phases, we can also differentiate
these explosions from the violent merger, spiral instability, and
detached DD models with the Co emission lines.

In addition, TD/QD explosions could be more luminous than these DD
models, if the companion WD is massive. We compare their luminosities,
assuming that their luminosities are proportional to the $^{56}$Ni
mass. TD and QD explosions yield $0.81$ and $1.01M_\odot$ of
$^{56}$Ni, respectively. On the other hand, the violent merger and
spiral instability models yield $\sim 0.6M_\odot$ of $^{56}$Ni mass at
most. $^{56}$Ni mass of the collisional DD model is at most
$0.4M_\odot$. The detached DD model yields $0.86M_\odot$ of
$^{56}$Ni. We find the TD and QD explosions are at least $1.4$ and
$1.7$ time more luminous than these DD models except the detached
model, respectively. The QD explosion is still $1.2$ times more
luminous than the detached DD model, while the TD explosion is as
luminous as the detached DD model. In summary, luminosity can be
useful to identify the TD/QD explosions from the other DD models,
except for the comparison between the TD explosion and detached DD
model.

\subsection{Detail Nucleosynthesis}

We derive detail nucleosynthesis using simulation data by a
post-processing method as follows. We pick up one SPH particle per
$4096$ SPH particles from three WDs: the primary WD in CO60He06 as a
representative of D$^6$ explosions, the companion WD in He45R09
(i.e. the companion WD of the TD explosion), and the companion WD in
CO90He54 (i.e. the companion WD in the QD explosion). Thus, the number
of SPH particles is $10240$ for the primary WD in CO60He06, $4608$ for
the companion WD in He45R09, and $9216$ for the companion WD in
CO90He54. We record density and temperature of these SPH particles
every timestep. Then, we calculate nucleosynthesis of these SPH
particles by detail nuclear reaction networks, using the {\tt torch}
code with $495$ nuclei \citep{1999ApJS..124..241T}. We adopt the solar
metallicity for the initial abundances of these SPH particles: 49.3~\%
C, 49.3~\% O, and 1.3~\% $^{22}$Ne in mass for SPH particles in
CO~cores, and 99~\% He and 1~\% $^{14}$N in mass for SPH particles in
He shells and He~WDs.

\begin{deluxetable}{l|lll}
  \tablecaption{Results of detail
    nucleosynthesis. \label{tab:DetailNucleosynthesis}}
  \tablehead{WDs & Fe & Mn/Fe & Ni/Fe \\
    & [$M_\odot$] & &
}
\startdata
pWD in CO60He06 & $0.59$ & $0.0025$ & $0.057$ \\
cWD in He45R09  & $0.55$ & $0.0031$ & $0.053$ \\
cWD in CO90He54 & $0.21$ & $0.0070$ & $0.15$  \\
\enddata
\tablecomments{The ``pWD'' and ``cWD'' indicates primary and companion
  WDs, respectively. The Mn/Fe and Ni/Fe shows their mass fractions.}
\end{deluxetable}

We show stable nuclei of Fe, Mn, and Ni for comparison with the
abundance pattern of 3C~397 in Table~\ref{tab:DetailNucleosynthesis}.
Note that we do not take into account SPH particles in the He shells
of the primary WD in CO60He06, and of the companion WD in CO90He54. We
compare these results with previous studies
\citep{2015ApJ...801L..31Y,2018ApJ...854...52S,2018ApJ...857...97M}. As
seen in the results of $1.0M_\odot$ CO~WD (i.e. the primary WD of
CO60He06), the mass fraction of Mn/Fe in our results is in good
agreement with the previous results, while the mass fraction of Ni/Fe
in our results is larger than in the previous results ($\sim
0.03$). This may come from our simple modeling of detonations.

Although the mass fractions of Ni/Fe in our simulations are larger
than those in previous studies, the mass fractions of Ni/Fe in D$^6$,
TD and QD explosions are much less than that of 3C~397 ($\gtrsim
0.1$). The mass fraction of Ni/Fe in the companion WD of He45R09 is
$0.15$. However, the total mass fraction of Ni/Fe in the TD explosion
is $0.081$. The mass fraction of Mn/Fe is also much less than that of
3C~397 ($\gtrsim 0.02$). Thus, D$^6$, TD, and QD explosions cannot
explain the abundance pattern of 3C~397. The progenitor of 3C~397 may
prefer a Chandrasekhar-mass explosion
\citep{2015ApJ...801L..31Y,2017ApJ...841...58D,2018ApJ...861..143L}.

\subsection{Hypervelocity White Dwarfs (HV~WDs)}

D$^6$ explosions form HV~WDs. Here, we consider the surface of HV~WDs
for HV~WD observations. Just after the explosions, the surface
materials consist of He, C, O, and $^{56}$Ni when the companion WD is
a CO~WD. The C/O materials come from the HV~WDs themselves. Since they
are marginally stripped by the explosions of the primary WDs, they can
be bound to the HV~WDs. The $^{56}$Ni materials are captured by the
HV~WDs from the supernova ejecta. The He materials come from the
primary WDs and He shells of HV~WDs themselves. Note that the
materials at the center of the primary WDs experience $\alpha$-rich
freezeout. If the companion WD is a He~WD, it has only He and
$^{56}$Ni on its surface.

It is not trivial that these chemical elements can be detected in
HV~WD observations. It should be clarified whether the captured
$^{56}$Ni materials and decay products stay on the surfaces of
HV~WDs. These $^{56}$Ni materials and decay products may settle down
into the inside of HV~WDs after some time
\citep{1986ApJS...61..197P,1992ApJS...82..505D}, or may stay on the
surface of HV~WDs due to radiative levitation
\citep{1995ApJS...99..189C,1995ApJ...454..429C}. The detection of
these He materials depends on the surface temperature of HV~WDs. Thus,
we should follow the long-term evolutions of HV~WDs in order to
directly compare simulation results of HV~WDs with HV~WD
observations. This is beyond the scope of this paper.

\section{Summary}
\label{sec:summary}

We have investigated nucleosynthesis signatures of stripping off and
inducing detonation of companion WDs in double-detonation explosions
for SNe~Ia. We have found various explosion types of these systems:
the D$^6$, TD, and QD explosions. The D$^6$ explosions occur in the
following conditions. The lighter WDs are CO~WDs with thin He shells,
or He~WDs separated from the heavier WDs by $\gtrsim
0.045R_\odot$. The D$^6$ explosions involve companion-origin
streams. The companion-origin streams contain C+O even when the
lighter WDs are CO~WDs with He shells, $\lesssim 0.04M_\odot$. These
materials contribute to low-velocity components in SNe~Ia. Thus, the
D$^6$ explosions can be counterparts of sub-luminous SNe~Ia, such as
iPTF14atg. Moreover, the companion-origin stream may contribute to
low-velocity C observed in several SNe~Ia.

The QD explosion arises if the lighter WD has a He shell, $\gtrsim
0.05M_\odot$. The QD explosion could be counterparts of luminous
SNe~Ia, such as SN~1991T and SN~1999aa, for the following two
reasons. (1) The QD explosion yields large amounts of $^{56}$Ni:
$1.01M_\odot$ for He90He54. (2) It should accompany early emissions
due to the He-detonation products, whose colors may be consistent with
the colors of SN~1991T-like and SN~1999aa-like SNe~Ia, such as
SN~2012cg and SN~2017cbv. If such QD explosions occur in DD systems
with less massive CO~WDs ($0.8$ -- $0.9M_\odot$), they can be
counterparts of SN~2007on with two-peak emission lines of Co~III.

The TD explosion may be achieved only when the primary WD has less
than $1.0M_\odot$, say $\lesssim 0.9M_\odot$. Then, the TD explosion
should yield $\lesssim 0.6M_\odot$ of $^{56}$Ni, which is not large
enough to explain luminous SNe~Ia. However, its spectra may be
consistent with luminous SNe~Ia, if it is observed from the side of
the companion WD.

The D$^6$ explosions can be identified from other DD models by oxygen
emission lines in the nebular phases, velocity shift, and signals of
He detonation products. The TD/QD explosions can be differentiated by
their luminosities and signals of He detonation products.

\acknowledgments

We thank Robert Fisher for fruitful advices. Numerical computations
were carried out on Oakforest-PACS at Joint Center for Advanced High
Performance Computing, and on Cray XC50 at Center for Computational
Astrophysics, National Astronomical Observatory of Japan. The software
used in this work was in part developed by the DOE NNSA-ASC OASCR
Flash Center at the University of Chicago.  This research has been
supported by World Premier International Research Center Initiative
(WPI Initiative), MEXT, Japan, by the Endowed Research Unit (Dark side
of the Universe) by Hamamatsu Photonics K.K., by MEXT program for the
Development and Improvement for the Next Generation Ultra High-Speed
Computer System under its Subsidies for Operating the Specific
Advanced Large Research Facilities, by ``Joint Usage/Research Center
for Interdisciplinary Large-scale Information Infrastructures'' and
``High Performance Computing Infrastructure'' in Japan (Project ID:
jh180021-NAJ, jh190021-NAJ), and by Grants-in-Aid for Scientific
Research (16K17656, 17K05382, 17H02864, 17H06360, 18H04585, 18H05223,
19K03907) from the Japan Society for the Promotion of Science.

\software{Modules of Helmholtz EoS and Aprox13 in FLASH
  \citep{2000ApJS..131..273F,2010ascl.soft10082F}}

\software{torch \citep{1999ApJS..124..241T}}

\appendix

In our simulations, we may highly simplify initiation and propagation
of detonations. In order to assess the simplifications, we compare
detonations in our simulations with those in more realistic
modelings. We investigate five detonations: He and CO detonations in
the primary WD of CO60He00 as representatives of those in the primary
WDs of all the models, He detonation in the companion WD of He45R09
(i.e. the third detonation in the TD explosion), and He and CO
detonations in the companion WD of CO90He54 (i.e. the third and fourth
detonations in the QD explosion, respectively).

In order to investigate detonations, we obtain physical quantities at
grid points we put optimally for each detonation. We use an SPH
interpolation to obtain these quantities, such that
\begin{align}
  q_{i} = \sum_j m_j \frac{q_j}{\rho_j}
  W\left(\left|\vec{r}_{i}-\vec{r}_j\right|,h_j\right),
\end{align}
where the subscripts $i$ and $j$ indicate quantities of a grid point
and a SPH particle, respectively, and $q$, $m$, $\rho$, $\vec{r}$, $h$
are general quantity, mass, density, position vector, and
kernel-support length, respectively. We adopt the $C^2$ Wendland
function for the SPH kernel function $W$
\citep{wendland1995piecewise,2012MNRAS.425.1068D} in the same way as
our simulations. The $C^2$ Wendland function in 3-dimension is
expressed as
\begin{align}
  W(r,h) = \frac{1}{h^3} (1-\hat{r})^4_+(1+4\hat{r}),
\end{align}
where $\hat{r}=r/h$, and $(\cdot)_+=\max(0,\cdot)$.

\begin{figure}[ht!]
  \plotone{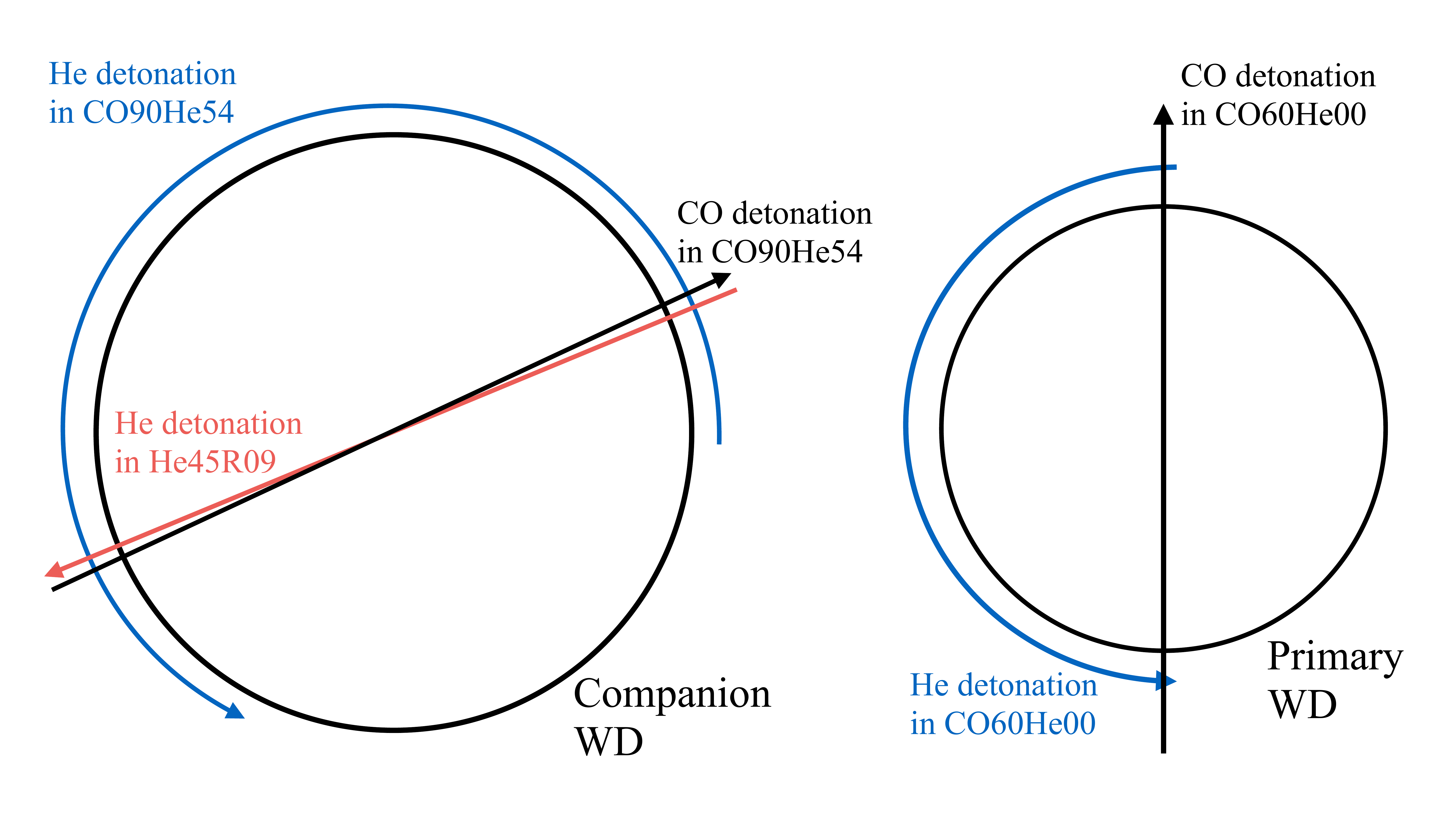}
  \caption{Lines and curves on which we put grid points in order to
    investigate the initiation and propagation of detonations. The
    detail definitions are described in Eqs.~(\ref{eq:HeCurveOfD6}) to
    (\ref{eq:COLineOfQD}).}
  \label{fig:lineOfDetonation}
\end{figure}

We put $1024$ uniformly spaced grid points along lines or curves for
investigating detonations (see
Figure~\ref{fig:lineOfDetonation}). These lines and curves can be
expressed as
\begin{align}
  \vec{r} &= \vec{r}_{\rm p} + R_{\rm p} \left( \cos (s/R_{\rm
    p}+\pi/2) \sin (s/R_{\rm p}+\pi/2), 0 \right) \;\; (0 \le s \le
  \pi R_{\rm p}) \label{eq:HeCurveOfD6} \\
  \vec{r} &= \vec{r}_{\rm c} + s \left( \cos (9\pi/8), \sin (9\pi/8),
  0 \right) \;\; (-L \le s \le L) \label{eq:HeLineOfTD} \\
  \vec{r} &= \vec{r}_{\rm c} + R_{\rm c} \left( \cos (s/R_{\rm c}),
  \sin (s/R_{\rm c}), 0 \right) \;\; (0 \le s \le 4\pi R_{\rm
    c}/3) \label{eq:HeCurveOfQD} \\
  \vec{r} &= \vec{r}_{\rm p} + s \left(0, 1, 0 \right) \;\; (-L \le s
  \le L) \label{eq:COLineOfD6} \\
  \vec{r} &= \vec{r}_{\rm c} + s \left( \cos (5\pi/36), \sin
  (5\pi/36), 0 \right) \;\; (-L \le s \le L) \label{eq:COLineOfQD}
\end{align}
for He detonation in the primary WD of CO60He00, He detonation in the
companion WD of He45R09, He detonation in the companion WD of
CO90He54, CO detonation in the primary WD of CO60He00, and CO
detonation in the companion WD of CO90He54, respectively. Here, $s$ is
the coordinate along the line or curve, and is used for the horizontal
axes of Figures~\ref{fig:HeliumHotspot} to
\ref{fig:CoreDetonation}. $\vec{r}_{\rm p}$ and $\vec{r}_{\rm c}$ are
the position vectors of the centers of the primary and companion WDs,
respectively. $R_{\rm p}$ and $R_{\rm c}$ are approximately the radii
at the bases of He shells of the primary and companion WDs,
respectively, where $R_{\rm p} = 4.6 \cdot 10^{3}$~km and $R_{\rm c} =
5.5 \cdot 10^3$~km. Finally, $L=10^4$~km.

In sections~\ref{sec:InitiationOfDetonation} and
\ref{sec:PropagationOfDetonation}, we show the initiation and
propagation of detonations, respectively.

\section{Initiation of detonation}
\label{sec:InitiationOfDetonation}

In this section, we confirm that we do not overproduce detonations.
We refer to \cite{2013ApJ...771...14H} (H13) and
\cite{2009ApJ...696..515S} (S09) as sufficient conditions of
initiation of He and CO detonations, respectively. We present He
detonations first, and CO detonations next.

The He detonation in CO60He00 starts from a hotspot we put
artificially. The hotspot has density of $\sim 10^6$~g~cm$^{-3}$,
temperature of $\sim 10^9$~K, and size of $\sim 10^3$~km (see
Figure~\ref{fig:initCO90He00}, for example). As for H13, such a
hotspot can start He detonation. Thus, we do not overproduce He
detonation in CO60He00. This is the same for the other models
undergoing D$^6$ and TD/QD explosions.

\begin{figure}[ht!]
  \plottwo{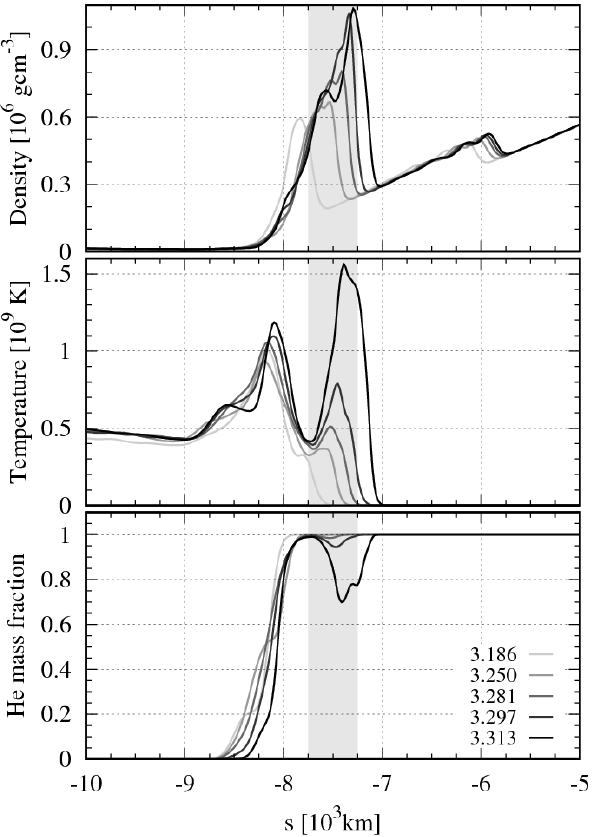}{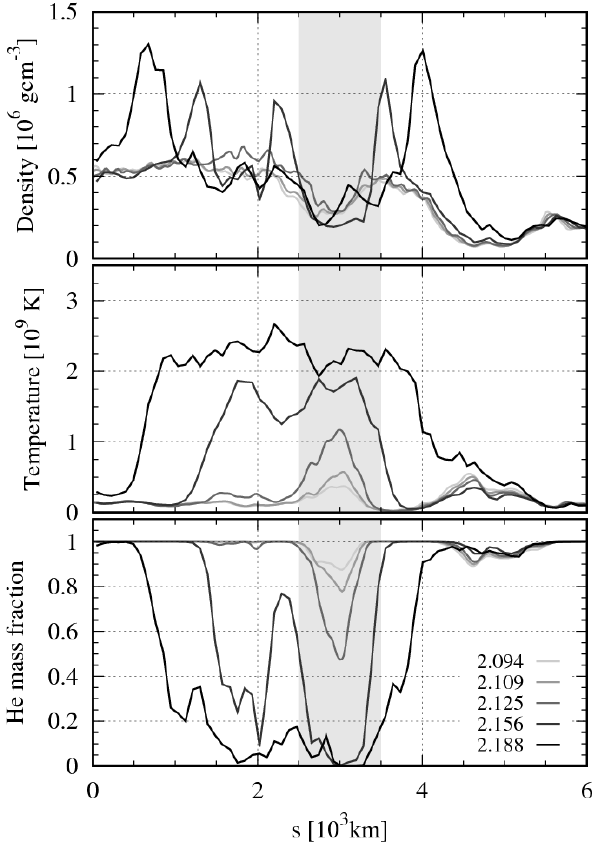}
  \caption{Formation of hotspots of He detonations in the companion
    WDs of He45R09 (left) and CO90He54 (right). The time is from
    $t=3.186$~s to $3.313$~s (left), and from $t=2.094$~s to $2.188$~s
    (right). The hotspot regions are indicated by shaded areas.}
  \label{fig:HeliumHotspot}
\end{figure}

Figure~\ref{fig:HeliumHotspot} shows the formation of hotspots of He
detonations in the companion WDs of He45R09 and CO90He54 as seen in
temperature evolution. The definition of $s$ in the horizontal axis
can be found in Eqs.~(\ref{eq:HeLineOfTD}) and (\ref{eq:HeCurveOfQD})
for the left and right panels, respectively.

We first explain properties of the hotspot of the He detonation in the
companion WD of He45R09. We can see the He mass fraction drastically
changes at $s \sim - 8 \cdot 10^3$~km, such that the He mass fraction
is zero at $s \lesssim -8.5 \cdot 10^3$~km, and unity at $s \gtrsim
-7.75 \cdot 10^3$~km. Thus, the surface of the companion WD is at $s
\sim - 8 \cdot 10^3$~km. Materials consist of supernova ejecta at $s
\lesssim - 8 \cdot 10^3$~km, and the companion WD at $s \gtrsim - 8
\cdot 10^3$~km. This is why the temperature at $s \lesssim - 8 \cdot
10^3$~km is high ($\gtrsim 5 \cdot 10^8$~K). There is a
high-temperature ($\sim 10^9$~K) region at $s \sim -8 \cdot
10^3$~km. At this region, the supernova ejecta collide with the
companion WD. The hotspot appears not at the center of the
high-temperature region, but at the skirt of this region inward the
companion WD at $s \sim -7.5 \cdot 10^3$~km. This may be because the
center of the high-temperature region has low density, $\lesssim
10^5$~g~cm$^{-3}$. The hotspot has density of $\sim 10^6$~g~cm$^{-3}$,
and size of $\sim 10^3$~km. At that density, a hotspot with size of
$\sim 10^2$~km can be large enough to ignite He detonation.

We next see the hotspot of the He detonation in the companion WD of
CO90He45. The hotspot appears at $s \sim 3 \cdot 10^3$~km. This is the
nearest side of the companion WD from the primary WD. The hotspot has
density of $3 \cdot 10^5$~g~cm$^{-3}$ and size of $\sim 10^3$~km. This
hotspot is also sufficiently large for He detonation.

\begin{deluxetable}{l|ccccc}
  \tablecaption{Hotspot sizes and space resolutions at the
    hotspots. \label{tab:Hotspot}}
  \tablehead{Model & CO06He00 (He) & He45R09 (He) & CO90He54 (He) &
    CO60He00 (CO) & CO90He54 (CO)}
\startdata
Size [km]       & $\sim 10^3$ & $\sim 10^3$ & $\sim 10^3$ & $\sim 10^2$ & $\sim 10^2$ \\
Resolution [km] & $36$        & $36$        & $45$        & $13$        & $13$ \\
\enddata
\tablecomments{The first line indicates model names and detonation
  types.}
\end{deluxetable}

These hotspots are well resolved in our simulations. We can calculate
effective space resolution of our simulations at these hotspots, using
Eq.~(\ref{eq:resolution}) in Section~\ref{sec:method}. In these
densities, the space resolutions are $\sim 36$~km for the former
($\sim 10^6$~g~cm$^{-3}$), and $\sim 45$~km for the latter ($\sim 5
\cdot 10^5$~g~cm$^{-3}$). These results are summarized in
Table~\ref{tab:Hotspot}.

\begin{figure}[ht!]
  \plottwo{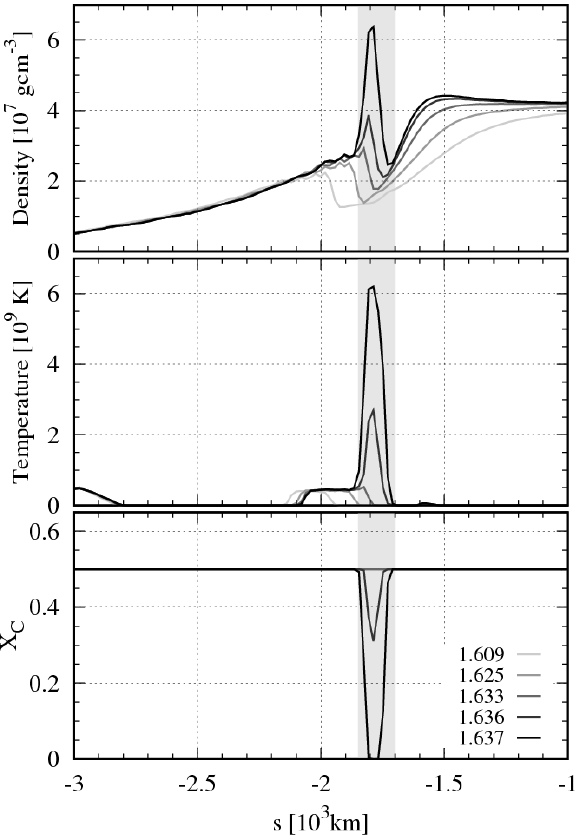}{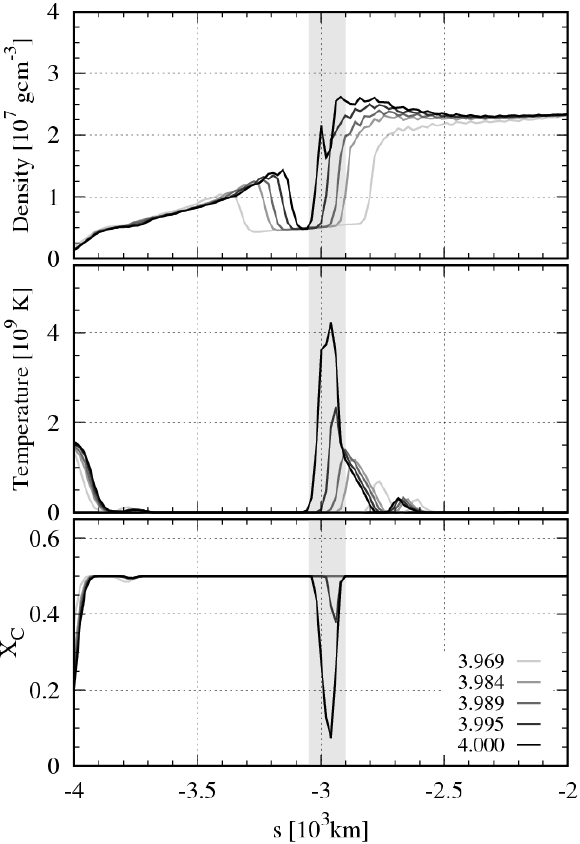}
  \caption{Formation of hotspots of CO detonations in the primary WD
    of CO60He00 (left), and in the companion WD of CO90He54
    (right). The time is from $t=1.609$~s to $1.637$~s (left), and
    from $t=3.969$~s to $4.000$~s (right). The hotspot regions are
    indicated by shaded areas.}
  \label{fig:CarbonHotspot}
\end{figure}

Figure~\ref{fig:CarbonHotspot} shows the formation of hotspots of CO
detonations in the primary WD of CO60He00, and in the companion WD of
CO90He54 as seen in the temperature evolution. The definition of $s$
in the horizontal axis can be found in Eqs.~(\ref{eq:COLineOfD6}) and
(\ref{eq:COLineOfQD}) for the left and right panels, respectively. For
both of the detonations, the hotspots have density of $\sim 2 \cdot
10^7$~g~cm$^{-3}$, and size of $\sim 10^2$~km. These conditions are
sufficient to generate CO detonation according to S09. Thus, these CO
detonations appropriately emerge.  These hotspots are also resolved in
our simulations. In these densities ($\sim 2 \cdot 10^7$~g~cm$^{-3}$),
the space resolutions are $\sim 13$~km for both the hotspots,
according to Eq.~(\ref{eq:resolution}) in Section~\ref{sec:method}.

As described above, we do not overproduce any He and CO
detonations. However, we may possibly miss the initiation of
detonation. In other words, we may fail to follow the initiation of He
and CO detonations from hotspots, even if the hotspots satisfy the
conditions of H13 and S09. Moreover, H13 and S09 have just shown
sufficient conditions for He and CO detonations,
respectively. \cite{2019ApJ...876...64F} have shown that turbulent
environment generates He and CO detonations more easily than predicted
by H13 and S09. Nevertheless, we do not miss the initiation of
detonations for all the models with respect of the conditions of H13
and S09. Moreover, turbulence are not effective in our setup, since we
do not follow mass transfer phases. In summary, we do not overproduce
nor miss any detonations in our simulations.

\section{Propagation of detonation}
\label{sec:PropagationOfDetonation}

In this section, we investigate detonation speeds, and jumps in
density and temperature due to the detonations. First, we show He
detonations in CO60He00, He45R09, and CO90He54, and next CO
detonations in CO60He00 and CO90He54.

\begin{figure}[ht!]
  \plotone{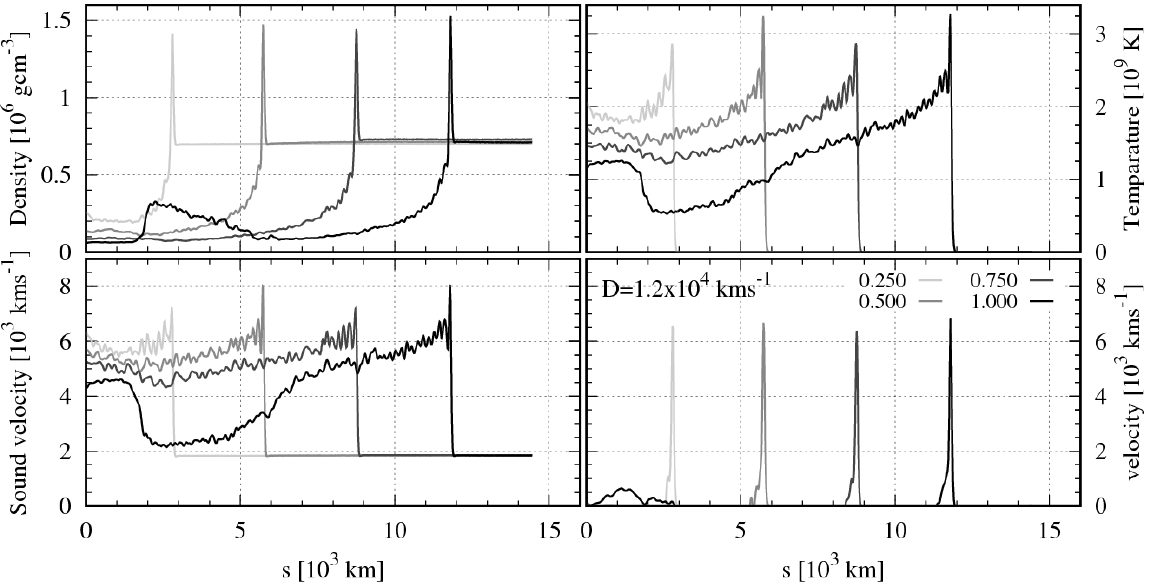}
  \caption{He detonation profiles in the primary WD for
    CO60He00. Density, temperature, sound velocity, and velocity at
    the rest frame of the center of the primary WD are shown from
    $t=0.25$~s to $t=1$~s. $D$ indicates the detonation speed relative
    to the fuel materials.}
  \label{fig:D6SurfaceDetonation}
\end{figure}

Figure~\ref{fig:D6SurfaceDetonation} shows the He detonation in the
primary WD of CO60He00. The definition of $s$ in the horizontal axis
can be found in Eq.~(\ref{eq:HeCurveOfD6}). The detonation speed is
$\sim 1.2 \cdot 10^4$~km~s$^{-1}$ at the rest frame of fuels. Since
the velocity of ashes is $\sim 5 \cdot 10^3$~km~s$^{-1}$ at the rest
frame of fuels, the ash velocity is $\sim 7 \cdot 10^3$~km~s$^{-1}$ at
the rest frame of the detonation, which is comparable to the sound
velocity of ashes. Thus, the detonation speed is consistent with the
Chapman-Jouguet (CJ) speed. The density jump is also consistent with
the CJ detonation as follow. In the strong limit of the CJ detonation,
the density ratio of the ash to the fuel should be $(\gamma_{\rm
  ash}+1)/\gamma_{\rm ash}$ according to Eq.~(129.15) of
\cite{1959flme.book.....L}, where $\gamma_{\rm ash}$ is the adiabatic
index of the ash. Since the ash pressure is dominated by radiation
pressure, $\gamma_{\rm ash} = 4/3$. Thus, the density ratio should be
$\lesssim 2$. The ash temperature is $\lesssim 3 \cdot 10^9$~K for the
fuel density with $7 \cdot 10^5$~g~cm$^{-3}$. This is compatible with
the ash temperature with $\gtrsim 3 \cdot 10^9$~K for the fuel density
with $10^6$~g~cm$^{-3}$ in H13.

\begin{figure}[ht!]
  \plotone{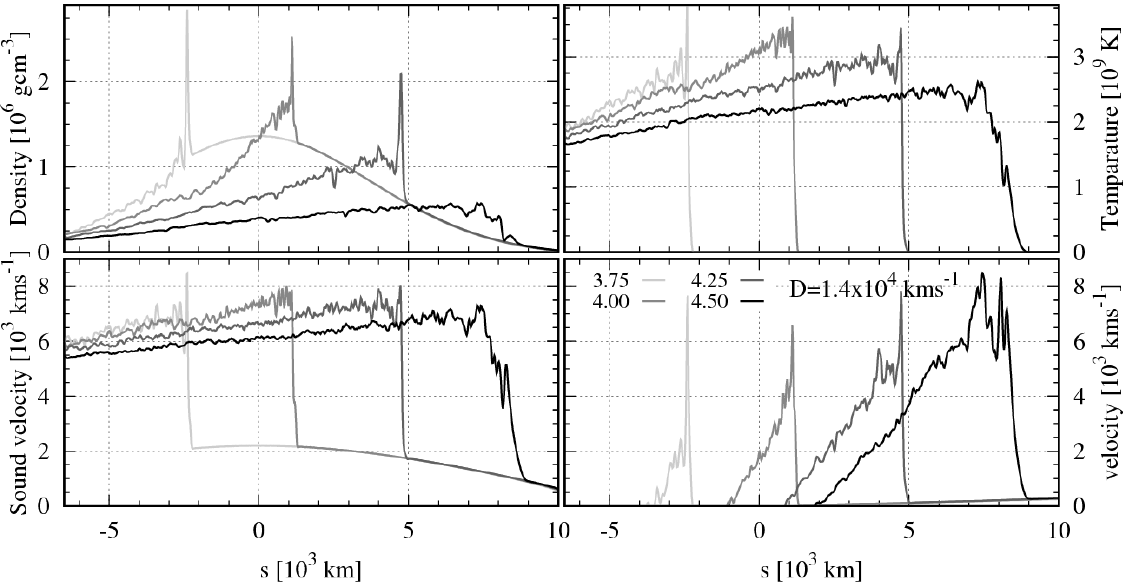}
  \caption{He detonation profiles in the companion WD for
    He45R09. Density, temperature, sound velocity, and velocity at the
    rest frame of the center of the companion WD are shown from
    $t=3.75$~s to $t=4.5$~s. $D$ indicates the detonation speed
    relative to the fuel materials.}
  \label{fig:TDCoreDetonation}
\end{figure}

Figure~\ref{fig:TDCoreDetonation} shows the He detonation in the
companion WD of He45R09. The definition of $s$ is described in
Eq.~(\ref{eq:HeLineOfTD}). We find the detonation speed is consistent
with the CJ speed, taking into account the ash velocity and the ash
sound velocity. Similarly to the He detonation in CO60He00, the
density jump is also consistent with that in the strong limit of the
CJ detonation. The ash temperature is $\sim 3 \cdot 10^9$~K for the
fuel density of $\sim 10^6$~g~cm$^{-3}$. This is in good agreement
with H13's results.

\begin{figure}[ht!]
  \plotone{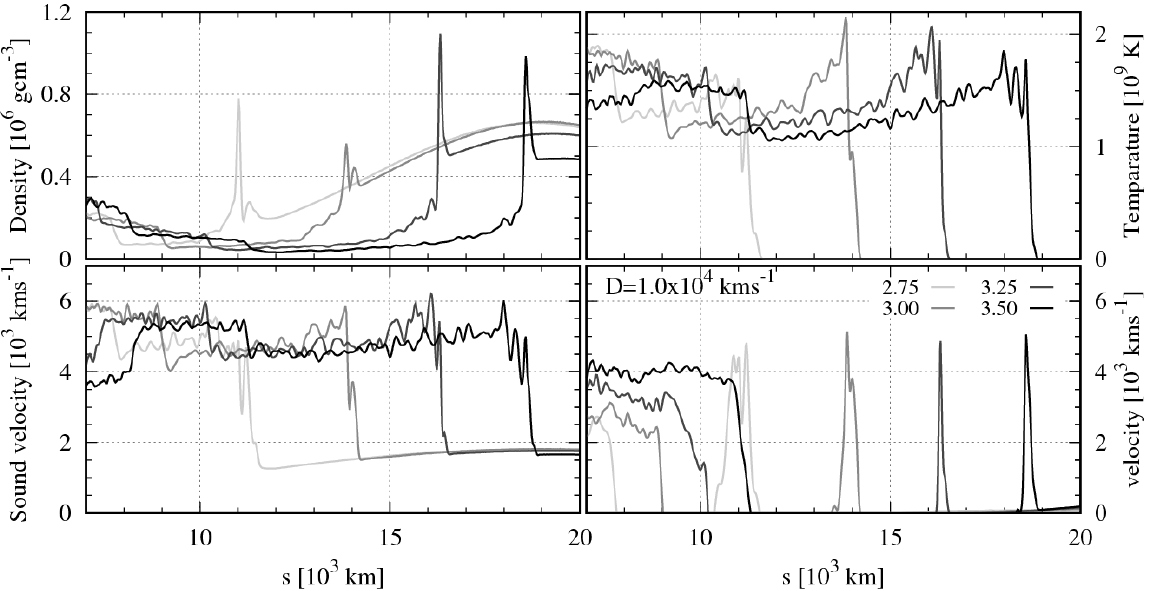}
  \caption{He detonation profiles in the companion WD for
    CO90He54. Density, temperature, sound velocity, and velocity at
    the rest frame of the center of the companion WD are shown from
    $t=2.75$~s to $t=3.5$~s. $D$ indicates the detonation speed
    relative to the fuel materials.}
  \label{fig:QDSurfaceDetonation}
\end{figure}

Figure~\ref{fig:QDSurfaceDetonation} shows the He detonation in the
companion WD of CO90He54. The horizontal axis $s$ is expressed in
Eq.~(\ref{eq:HeCurveOfQD}). The detonation speed is $\sim 1.0 \cdot
10^4$~km~s$^{-1}$, consistent with the CJ speed. The density jump is
also consistent with that in the strong limit of the CJ
detonation. The ash temperature is $\sim 2 \cdot 10^9$~K for the fuel
density of $< 6 \cdot 10^5$~g~cm$^{-3}$. This is compatible to H13's
results.

\begin{figure}[ht!]
  \plottwo{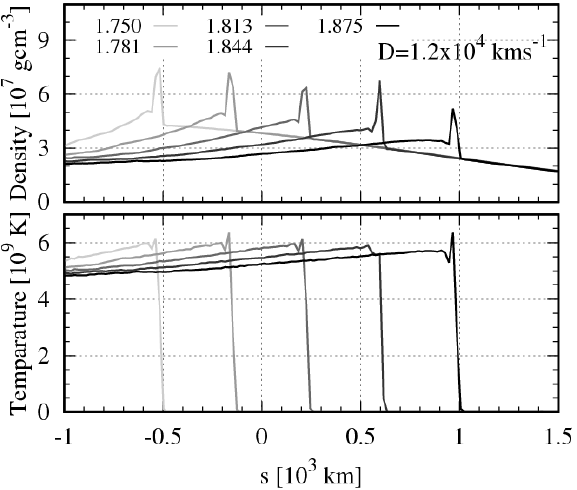}{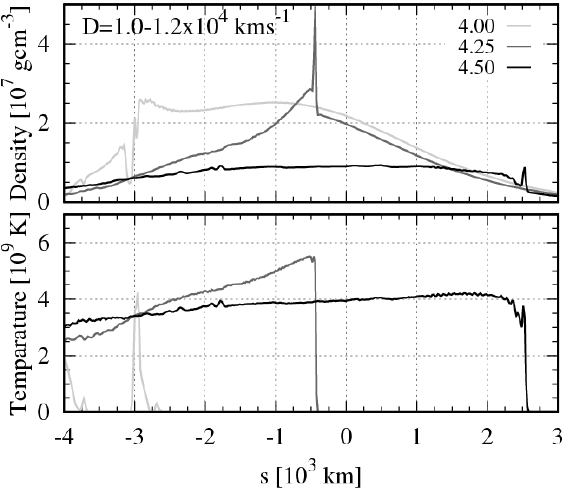}
  \caption{CO detonations in the primary WD of CO60He00 (left), and in
    the companion WD of CO90He54 (right). $D$ indicates the detonation
    speed relative to the fuel materials.}
  \label{fig:CoreDetonation}
\end{figure}

Figure~\ref{fig:CoreDetonation} shows the CO detonations in the
primary WD of CO60He00, and in the companion WD of CO90He54. The
definitions of $s$ in the left and right panels are the same as the
left and right panels in Figure~\ref{fig:CarbonHotspot},
respectively. Since the fuel density exceeds $2 \cdot
10^7$~g~cm$^{-3}$ for both the CO detonations, these detonations
should be so-called a pathological type of detonation
\citep{1989MNRAS.239..785K}. According to \cite{1999ApJ...512..827G},
such detonations have a speed of $\sim 1.2 \cdot 10^4$~km~s$^{-1}$
when the fuel density is $2$ - $3 \cdot 10^7$~g~cm$^{-3}$. This is
consistent with the CO detonations in our models. Moreover,
\cite{1999MNRAS.310.1039S} have shown that the ash density and
temperature are, respectively, $\sim 7 \cdot 10^7$~g~cm$^{-3}$ and $6
\cdot 10^9$~K in the fuel density of $\sim 4 \cdot
10^7$~g~cm$^{-3}$. This is in good agreement with the CO detonation at
$t=1.75$~s in CO60He00.


\end{document}